\documentclass[referee,a4paper,12pt,traditabstract]{jswsc} 

\usepackage{amsmath}
\usepackage{physics}
\usepackage{graphicx}
\usepackage{xcolor}
\usepackage{txfonts}
\usepackage{subcaption}
\usepackage{epstopdf}
\usepackage[displaymath,mathlines]{lineno}
\usepackage[authoryear,round]{natbib}
\usepackage[backref]{hyperref}
\usepackage{url}

\newcommand{\diff}{\mathrm{d}}
\newcommand{\vect}[1]{\boldsymbol{#1}}

\hypersetup{colorlinks=true,citecolor=cyan,urlcolor=cyan,linkcolor=blue}

\begin{document}


   \title{Energetic particle acceleration and transport with the novel Icarus+PARADISE model}

   \titlerunning{Icarus$+$PARADISE model}

   \authorrunning{Husidic et al.}

   \author{E. Husidic
          \inst{1,2}
          \and
          N. Wijsen
          \inst{1}
          \and
          T. Baratashvili
          \inst{1}
          \and
          S. Poedts
          \inst{1,3}
          \and
          R. Vainio
          \inst{2}
          }

   \institute{Centre for mathematical Plasma Astrophysics, Department of Mathematics,\\ KU Leuven, Celestijnenlaan 200B, 3001 Leuven, Belgium\\
              \email{\href{mailto:edin.husidic@kuleuven.be}{edin.husidic@kuleuven.be}}
         \and
             Department of Physics and Astronomy, University of Turku, 20014 Turku, Finland
        \and
            Institute of Physics, University of Maria Curie-Sk{\l}odowska, ul.\ Radziszewskiego 10, \\ 
            20-031 Lublin, Poland
             }

   \date{Received October 02, 2023; accepted March 26, 2024}

  \abstract 
   {
     With the rise of satellites and mankind's growing dependence on technology, there is an increasing awareness of space weather phenomena related to high-energy particles. Shock waves driven by coronal mass ejections (CMEs) and corotating interaction regions (CIRs) occasionally act as potent particle accelerators, generating hazardous solar energetic particles (SEPs) that pose risks to satellite electronics and astronauts. Numerical simulation tools capable of modelling and predicting large SEP events are thus highly demanded. We introduce the new Icarus$+$PARADISE model as an advancement of the previous EUHFORIA$+$PARADISE model. Icarus, based on the MPI-AMRVAC framework, is a three-dimensional magnetohydrodynamic code that models solar wind configurations from 0.1~au onwards, encompassing transient structures like CMEs or CIRs. Differing from EUHFORIA's uniform-only grid, Icarus incorporates solution adaptive mesh refinement (AMR) and grid stretching. The particle transport code PARADISE propagates energetic particles as test particles through these solar wind configurations by solving the focused transport equation in a stochastic manner. We validate our new model by reproducing EUHFORIA$+$PARADISE results. This is done by modelling the acceleration and transport of energetic particles in a synthetic solar wind configuration containing an embedded CIR. Subsequently, we illustrate how the simulation results vary with grid resolution by employing different levels of AMR. The resulting intensity profiles illustrate increased particle acceleration with higher levels of AMR in the shock region, better capturing the effects of the shock. 
   }


   \maketitle

\section{Introduction}\label{sec:introduction}

    Space weather events are having an ever growing socio-economic impact on today's societies that are heavily reliant on nanotechnology, Global Navigation Satellite Systems (GNSS), and satellites for communication \citep{Hapgood-2011, Schrijver-etal-2015}. Solar eruptive events, especially coronal mass ejections (CMEs), stand out as the primary drivers of hazardous conditions in the inner heliosphere that spacecraft and astronauts are exposed to \citep{Gopalswamy-2018}. Moreover, fast CMEs directed towards Earth  can interact with the terrestrial magnetic field, causing geomagnetic storms and potentially affecting electrical power grids, pipelines, and telecommunication \citep{Hapgood-2011, Schrijver-etal-2014, Schrijver-etal-2015}.

    A frequent consequence of solar eruptions is the acceleration of charged particles that are collectively referred to as solar energetic particles (SEPs), comprising electrons, protons, and ions \citep{Reames-2017}. 
    Historically, two types of SEP events were distinguished, namely impulsive and gradual SEP events, where the first type was linked to solar flares and the second type to CMEs \citep{Reames-1999}. However, \citet{Kallenrode-2003}, and later also \citet{Cane-etal-2010}, showed that the largest SEP events include contributions from both flare and CME shock acceleration. Beside solar eruptive phenomena, the shocks in corotating interaction regions (CIRs) function occasionally as particle accelerators to generate energetic particles in the inner heliosphere, and also accelerate more efficiently particles in the outer heliosphere \citep{Richardson-2004, Wijsen-etal-2019}. A CIR is a region where a fast solar wind stream catches up a slow solar wind stream and compresses it. Some authors \citep{Jian-etal-2006,Richardson-2018} reserve the term CIR for interaction regions with a lifetime greater than one solar rotation, and refer to events of shorter duration as stream interaction regions (SIRs). In the present work we refer to all such interaction regions as CIRs. At radial distances well beyond 1 au, CIRs  are often bounded by two shock waves, a forward shock wave accelerating the slow wind, and a reverse shock decelerating the fast wind. However, in the majority of cases observed at 1~au, CIRs are only bounded by two compression waves, and the shock waves form further out \citep{Richardson-2018}. In observational data, the suprathermal particle intensity profiles often show a double-peaked profile as a result of particles being accelerated at both the forward and reverse shock/compression waves \citep{Smith-Wolfe-1976, Richardson-2018, Allen-etal-2021,Wijsen-etal-2021}.

     In large SEP events, accelerated particles can achieve energies of several hundreds of MeV/nuc or even up to a few GeV/nuc, posing a dangerous threat to satellites and astronauts. Therefore, there is significant interest in understanding the acceleration and transport processes behind SEP events. Despite huge and decades-long efforts, there is still no consensus in the scientific community about the exact acceleration mechanisms for SEPs \citep{Petrosian-2016, Perri-etal-2022}. A frequently proposed candidate for the acceleration mechanism in impulsive events at solar flare sites during magnetic reconnection is stochastic acceleration (SA) due to turbulence \citep{Petrosian-Liu-2004}. Possible explanations for the acceleration of SEPs at shock fronts include diffusive shock acceleration (DSA) and shock drift acceleration (SDA), accompanied by SA \citep{Desai-Giacalone-2016, Vainio-Afanasiev-2017}. In the SDA mechanism that occurs when the charged particle interacts with a quasi-perpendicular shock front, the particle travels multiple times upstream and downstream the shock. The difference in magnetic field strengths in the two regions leads to a change in the cyclotron radius of the particle which shift the particle's guiding centre along the electric field, resulting in a continuous acceleration. The DSA mechanism dominates at quasi-parallel shocks. Here, the charged particle also crosses multiple times the shock front to gain energy, but is scattered at frozen-in magnetic turbulence on both sides of the shock, altering the particle's pitch angle. The SA occurs due to resonant wave-particle interactions between the SEP and magnetic field fluctuations (for more details on the acceleration mechanisms, see e.g., \citealt{Kallenrode-2004, Vainio-Afanasiev-2017}). 

    In recent years, several simulation models have been developed both for theoretical studies and the forecasting of SEP and CIR events, characterised by a common feature, the coupling of a heliospheric solar wind model with a particle transport code. A few of these models shall be mentioned in the following. The SEPMOD (Solar Energetic Particle MODel) developed by \citet{Luhmann-etal-2017}  uses background solar wind configurations generated by the heliospheric three-dimensional (3D) magnetohydrodynamic (MHD) model Enlil \citep{Odstrcil-etal-2004, Odstrcil-etal-2005} to simulate the scatter-free transport of SEPs. Another SEP propagation model is the Energetic Particle Radiation Environment Module (EPREM), which includes diffusion and drift effects of the particles, introduced by \citet{Kozarev-etal-2010}. EPREM was initially coupled to the ENLIL model \citep{Kozarev-etal-2010}, while later \citet{Kozarev-etal-2013} advanced EPREM to take input from the Block Adaptive Tree Solar-Wind Roe Upwind Scheme (BATS-R-US) model \citep{Manchester-etal-2012}. Recently, the PArticle Radiation Asset Directed at Interplanetary Space
    Exploration (PARADISE) code \citep{Wijsen-2020} has been introduced that uses solar wind background configurations by the heliospheric 3D MHD model EUropean Heliospheric FORecasting Information Asset (EUHFORIA; \citealt{Pomoell-Poedts-2018}) to propagate SEPs as test particles and calculate the intensity distributions of the energetic particles. Similar to ENLIL, EUHFORIA incorporates a uniform equidistant grid. EUHFORIA$+$PARADISE has been applied successfully both in theoretical studies with synthetic solar wind maps \citep{Wijsen-etal-2019} and by reproducing real SEP events \citep{Wijsen-etal-2021, Wijsen-etal-2023}.

    In the present work, we introduce our new model, Icarus$+$PARADISE. Similar to EUHFORIA, Icarus is a 3D MHD heliospheric wind model that simulates solar wind configurations from 0.1~au onward, and models the propagation and evolution of transient structures such as CMEs and CIRs \citep{Verbeke-etal-2022}. When Icarus uses an equidistant grid, it functions similarly to EUHFORIA or ENLIL. However, unlike the latter models, Icarus can incorporate advanced numerical techniques such as radial grid stretching and (solution) adaptive mesh refinement (AMR), to increase or lower the spatial resolution in selected regions of the computational domain. For demonstration purposes, we choose in this work a synthetic idealised solar wind map to provide the inner boundary conditions at 0.1~au. In the resulting heliospheric solar wind model that contains CIRs, we use our particle transport code PARADISE to inject 1~MeV protons in the forward and reverse shock of the CIR. By solving the focused transport equation (FTE; see e.g., \citealt{vandenBerg-etal-2020} and references therein), intensities of the energetic particles are obtained. To validate our new model, we compare simulation results to the EUHFORIA$+$PARADISE model that has been successfully tested with real data. Finally, we apply different levels of AMR to the background solar wind configurations of Icarus and show the significance of an increased resolution at the shock waves for the obtained particle intensities.

   The paper is structured as follows. In Sec.~\ref{sec:models_and_setup} we describe the energetic particle code PARADISE (Sec.~\ref{subsec:paradise}), the
   heliospheric models Icarus and EUHFORIA (Sec.~\ref{subsec:heliospheric_models}), the setup for the subsequent simulations (Sec.~\ref{subsec:setup}) as well as considerations made for the coupling of PARADISE to Icarus (Sec.~\ref{subsec:coupling_PARADISE_Icarus}). Results of the simulations are displayed in Sec.~\ref{sec:results}, where the Icarus$+$PARADISE model is 
   validated by reproducing results of the established EUHFORIA$+$PARADISE model 
   (Sec.~\ref{subsec:validation}), and subsequently applied with different levels of AMR
   (Sec.~\ref{subsec:amr_runs}). The paper concludes with a summary and discussion in 
   Sec.~\ref{sec:summary_conclusions}.

\section{Numerical models and setup}\label{sec:models_and_setup}

    The Icarus$+$PARADISE model represents a coupling of two models. The PARADISE code is used to calculate the spatio-temporal evolution of energetic particles, and its functionality and setup are described in Sec.~\ref{subsec:paradise}. PARADISE requires as input background solar wind configurations which are provided by the Icarus code. Together with the previously coupled EUHFORIA model that is used here to validate the Icarus$+$PARADISE, the two heliospheric models are illustrated in Sec.~\ref{subsec:heliospheric_models}. The setup for the simulations in Sec.~\ref{sec:results} is described in Sec.~\ref{subsec:setup}, while Sec.~\ref{subsec:coupling_PARADISE_Icarus} outlines considerations and challenges regarding the coupling of PARADISE to Icarus.

   \subsection{The PARADISE code}\label{subsec:paradise}

    PARADISE is a particle transport code utilised for solving the focused transport equation (FTE) to derive spatio-temporal distributions of energetic particles \citep{Wijsen-2020}. The FTE \citep{Roelof-1969,Skilling-1971,Skilling-1975, Ruffolo-1995, Isenberg-1997, le-Roux-Webb-2009} can be derived from the Vlasov equation under the assumption that the energetic particles propagate as test particles through a prescribed solar wind background. It describes the evolution of the five-dimensional gyrotropic particle distribution function $f(\vect{x},p,\mu,t)$, which depends on various parameters: spatial coordinates $\vect{x}$, momentum magnitude $p$, pitch angle cosine $\mu \equiv \cos (\alpha)$ with pitch angle $\alpha = \arctan (v_\perp / v_\parallel)$ where, respectively, $v_\perp$ and  $v_\parallel$ denote the velocity components perpendicular and parallel with respect to the ambient mean magnetic field, and time $t$. PARADISE implements the FTE in the form
   \begin{align}
       \frac{\partial f}{\partial t} 
       + \frac{\diff\vect{x}}{\diff t} \cdot \nabla f 
       + \frac{\diff \mu}{\diff t} \frac{\partial f}{\partial \mu} 
       + \frac{\diff p}{\diff t} \frac{\partial f}{\partial p}
       = \frac{\partial}{\partial \mu} \left(D_{\mu\mu} \frac{\partial f}{\partial \mu} \right)
       + \nabla \cdot \left( \vect{\kappa}_\perp \cdot \nabla f \right) \label{eq:FTE}
   \end{align}
   with $D_{\mu\mu}$ being the pitch angle diffusion coefficient, $\vect{\kappa}_\perp$
   the spatial cross-field diffusion tensor (see below for details). 
   The total derivatives of $\vect{x}$, $\mu$, and $p$
   in Eq.~\eqref{eq:FTE} are given by
    \begin{align}
        \frac{\diff \vect{x}}{\diff t} 
        &= \vect{V}_\mathrm{sw} + \vect{V}_\mathrm{d} + \mu\,v\,\vect{b} 
        \label{eq:dxdt}\,, \\
        \frac{\diff \mu}{\diff t} 
        &= \frac{1 - \mu^2}{2} \left(v\,\nabla \cdot \vect{b} 
        + \mu\,\nabla \cdot \vect{V}_\mathrm{sw}
        - 3\,\mu\,\vect{b}\vect{b}:\nabla \vect{V}_\mathrm{sw}
        - \frac{2}{v}\vect{b} \cdot \frac{\diff \vect{V}_\mathrm{sw}}{\diff t}\right)\,,
        \label{eq:dmudt} \\
        \frac{\diff p}{\diff t} &= 
        \left[\frac{1 - 3\,\mu^2}{2} \left(\vect{b}\vect{b}:\nabla \vect{V}_\mathrm{sw}\right)
        - \frac{1-\mu^2}{2} \nabla \cdot \vect{V}_\mathrm{sw}
        - \frac{\mu}{v} \vect{b} \cdot \frac{\diff \vect{V}_\mathrm{sw}}{\diff t}\right]p\,.
        \label{eq:dpdt}
    \end{align}
    Here, $\vect{V}_\mathrm{sw}$ denotes the solar wind bulk velocity, $\vect{V}_\mathrm{d}$ the
    drift velocity of a particle's guiding center (GC) accounting for magnetic drifts (in the present work we are not concerned with GC drifts and thus set $\vect{V}_\mathrm{d} = \vect{0}$), $v \equiv p/(\gamma\,m)$ is the particle
    speed with $\gamma$ being the Lorentz factor and $m$ the particle rest mass, and $\vect{b}$ the
    unit vector in direction of the mean magnetic field. Furthermore, the colon in Eqs.~\eqref{eq:dmudt} and \eqref{eq:dpdt} denotes the Frobenius inner product, which can be expressed by using Einstein's summation convention as $\vect{b}\vect{b}:\nabla \vect{V}_\mathrm{sw} = b_{ij}\, \partial v_i / \partial x_j$. The FTE in Eqs.~\eqref{eq:FTE} to \eqref{eq:dpdt} is written in mixed coordinates, that is, the spatial coordinates are measured in an observer's inertial frame, while the momentum magnitude and the pitch angle are given in a frame co-moving with the solar wind. 
    We note that Eqs.~\eqref{eq:FTE} -- \eqref {eq:dpdt} ignore terms of the order of ${V}_\mathrm{sw} / c$, where $c$ denotes the speed of light. This is a reasonable approximation, given that ${V}_\mathrm{sw}/c \sim 10^{-3}$. For a derivation of the FTE retaining terms of higher order in ${V}_\mathrm{sw} / c$, see for instance, Ch.~2 of \citet{Wijsen-2020}, and references therein.
    Inside PARADISE, Eqs.~\eqref{eq:FTE} to \eqref{eq:dpdt} are rewritten to an equivalent set of stochastic differential equations (SDEs) that are integrated forward in time using It$\hat{\mathrm{o}}$ calculus \citep{Wijsen-2020, Strauss-Effenberger-2017}.

    Global heliospheric MHD models, such as EUHFORIA and Icarus (see Sec.~\ref{subsec:heliospheric_models}), derive large-scale solar wind configurations, where the interplanetary magnetic field (IMF) follows a nominal spiral pattern unless it is perturbed by transient structures like CMEs and CIRs. However, even in the absence of such structures, on small spatial and temporal scales, the IMF exhibits turbulent fluctuations that are not captured by the MHD model and which can, for instance, scatter the energetic particles. In the FTE, the effect of such turbulent fluctuations on the energetic particle transport is modelled through a set of diffusion processes in phase space. An expression of these diffusion coefficients can be derived from, for instance, quasi-linear theory \citep{Jokipii-1966}.
    
    In PARADISE, the diffusion processes are treated in a modular fashion, that is, the diffusion terms in the FTE describing turbulence can be added, removed, or exchanged. 
    PARADISE contains different expressions for the pitch angle diffusion coefficient which are derived under the assumption of quasi-linear theory (QLT) and magnetic slab turbulence (see e.g., \citealt{Jokipii-1966} or \citealt{Jackel-Schlickeiser-1992}). 
    In the present work, we use an expression similar to those of \citet{Agueda-etal-2008} and \citet{Agueda-Vainio-2013}, given by
    \begin{align}
        D_{\mu\mu} = D_0 \left(\frac{\vert \mu \vert}{1 + \vert \mu \vert} + \epsilon \right) \left(1 - \mu^2\right)\,.\label{eq:Dmumu}
    \end{align}
    Here, $\epsilon$ describes a parameter to resolve the resonance gap at $\mu = 0$ (see e.g., \citealt{Klimas-Sandri-1971}; in the present work, $\epsilon = 0.048$), and the scaling factor $D_0$ is determined by the particle's rigidity and assuming a parallel mean free path (MFP) that is constant in space. In particular,  we take the parallel MFP as $\lambda_\parallel = 0.3$~au for a 1 MeV proton. For details, we refer to \citet{Wijsen-2020}.

    Beside pitch angle diffusion, we consider also cross-field spatial diffusion (CFD) described by the diffusion tensor \textbf{\textit{$\kappa_\perp$}}. Spatial CFD is a proposed mechanism to explain, for instance,  SEP events that are observed by spacecraft which are seemingly  not magnetically connected to the particle source \citep[e.g.,][]{Klassen-etal-2015}. 
    In this work, we account for CFD in the PARADISE simulations by prescribing a constant perpendicular MFP length of $\lambda_\perp = 3\times 10^{-4}$~au. Thus, with $\lambda_\perp/\lambda_\parallel = 10^{-3}$, the particles will predominantly propagate along the IMF lines in our simulations.

   \subsection{Heliospheric solar wind models}\label{subsec:heliospheric_models}

    Icarus, as detailed in \citet{Verbeke-etal-2022} and \citet{Baratashvili-etal-2022}, operates within a reference frame corotating with the Sun. Icarus solves the ideal 3D MHD equations to obtain heliospheric solar wind configurations from 0.1~au onward. A simple Icarus solar wind and a cone CME model is maintained as a test case in the Message Passing Interface - Adaptive Mesh Refinement Versatile Advection Code (MPI-AMRVAC), which constitutes a collection of various numerical schemes tailored for handling (near-)conservation laws and shock problems in hydrodynamics and magnetohydrodynamics \citep{Keppens-etal-2012, Xia-etal-2018}. One notable feature of Icarus is its flexibility regarding the choice of a coronal model that provides the inner boundary conditions, as long as the solar wind is considered super-Alfv\'{e}nic at 0.1~au (in fact, the speed has to be above the fast magnetosonic speed). While this is a common approach in MHD solar wind modelling, recent observations by the Parker Solar Probe indicate that the solar wind can still be sub-Alfv\'{e}nic at distances beyond 0.1~au \citep{Bandyopadhyay-etal-2022, Jiao-etal-2024}. Moreover, Icarus is able to exploit the output of EUHFORIA's coronal model (see next paragraph), providing it with the inner boundary conditions. The time advantage of radial grid stretching and AMR applications in Icarus compared to simulations of EUHFORIA employing an equidistant grid has been demonstrated in studies conducted by \citet{Verbeke-etal-2022} and \citet{Baratashvili-etal-2022}.

\begin{figure*} [h!]
\centering
    \subfloat{{\includegraphics[scale=0.21]{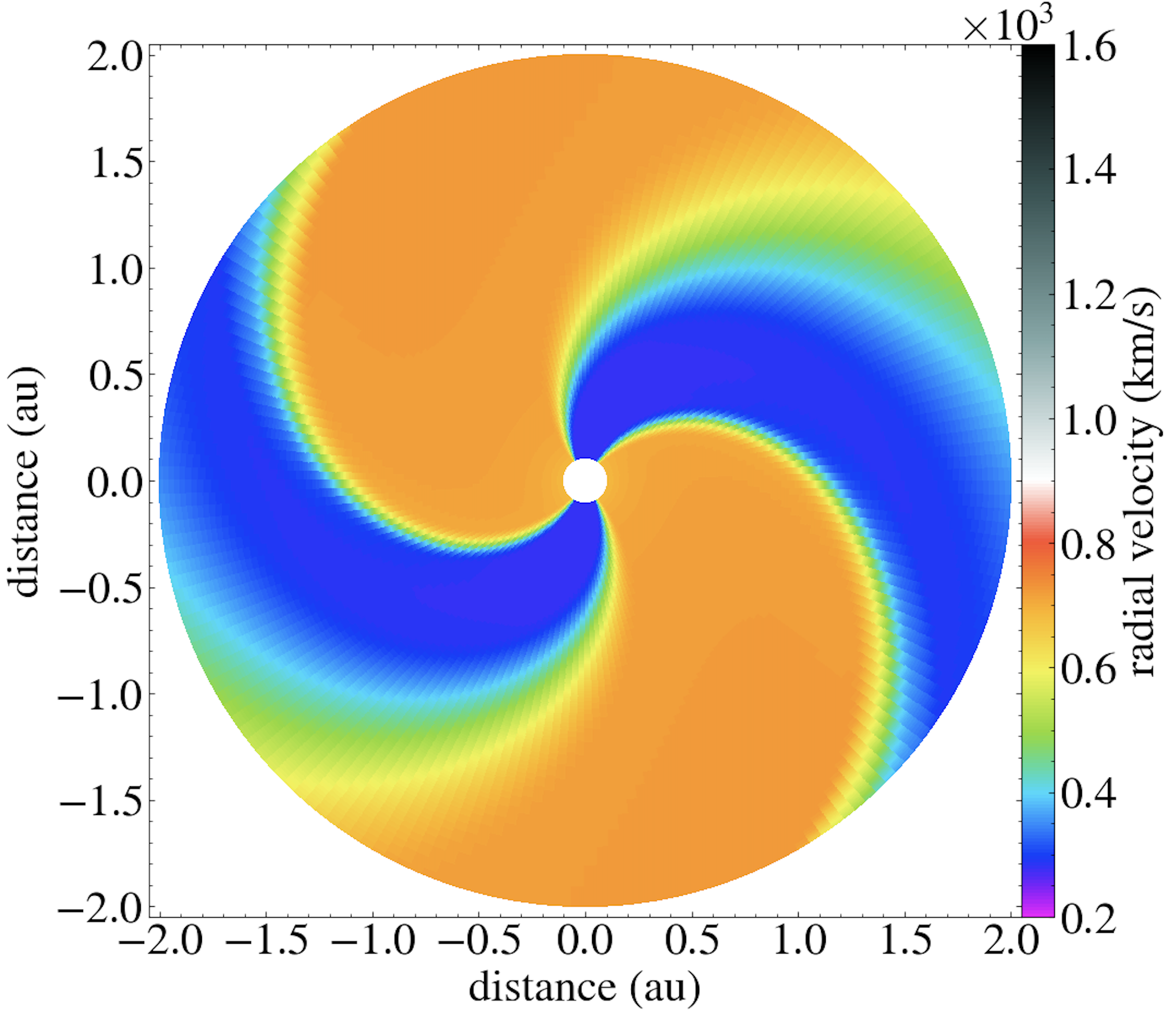} }}%
    \qquad
    \subfloat{{\includegraphics[scale=0.21]{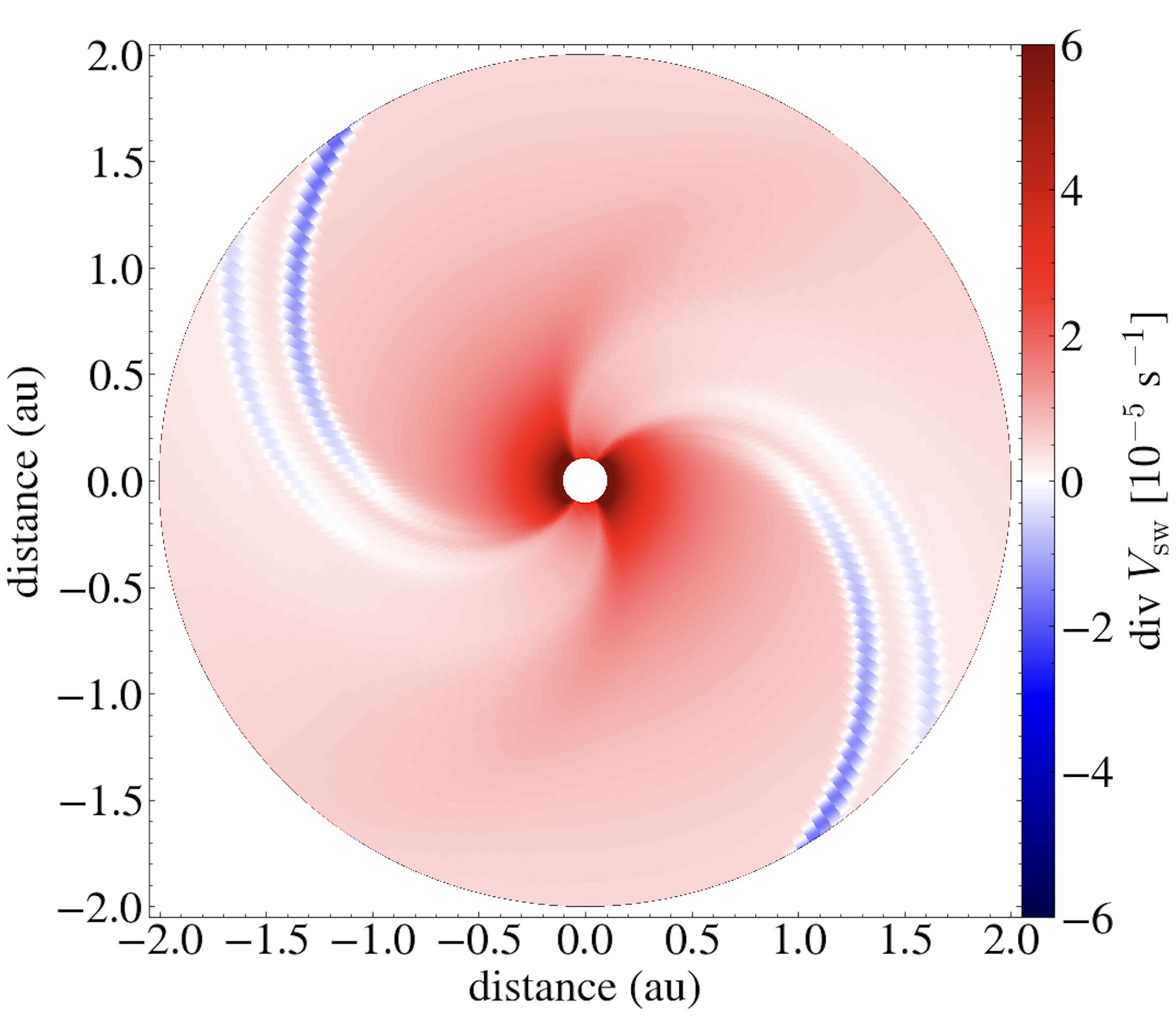} }}%
    \caption{Contour plots showing the radial velocity component in the ecliptic plane from 0.1~au to 2~au by Icarus (left panel), and the corresponding divergence of the solar wind velocity in the Icarus simulation.}%
    \label{fig:vr_comparison}%
\end{figure*}

    EUHFORIA is a physics-based space weather forecasting tool that generates solar wind configurations similar to Icarus from 0.1~au onward, including transient structures such as CMEs \citep{Pomoell-Poedts-2018}. The model comprises two primary components, a coronal and a heliospheric part. The semi-empirical coronal model accepts as input synoptic magnetograms by the Global Oscillation Network Group (GONG) and employs the Wang-Sheeley-Arge model in order to obtain the solar wind parameters at the inner boundary. Using these solar wind parameters as inner boundary conditions, the heliospheric part of EUHFORIA solves the ideal 3D MHD equations to generate solar wind configurations from 0.1~au onwards. Unlike Icarus, EUHFORIA solves the ideal MHD equations using Heliocentric Earth EQuatiorial (HEEQ) coordinates with Earth's longitude fixed at $0^\circ$ on an equidistant grid without the ability of AMR.

       \begin{figure}
   \centering
   \includegraphics[scale=0.25]{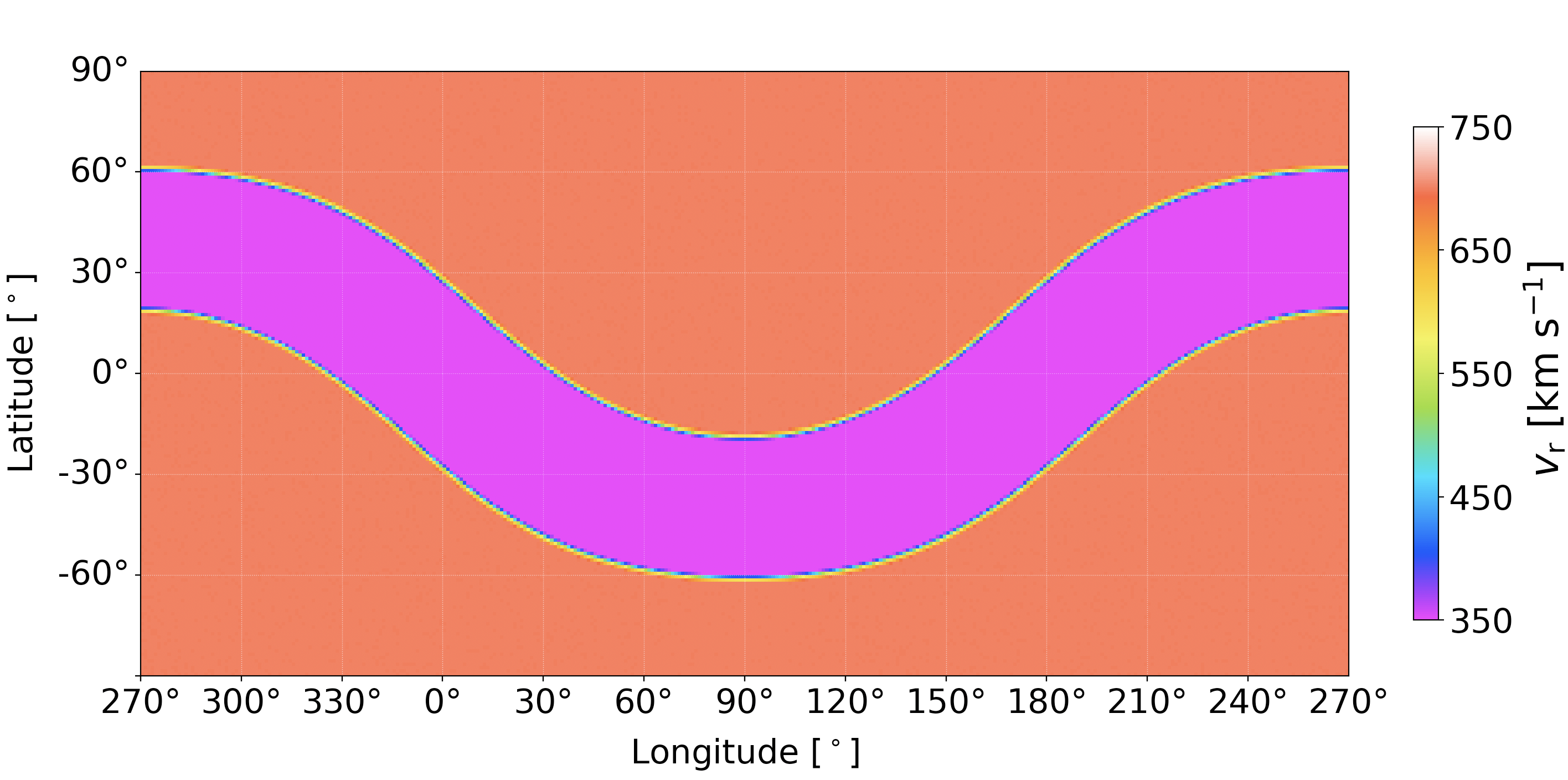}
      \caption{Contour plot of the radial speed at the inner boundary both for Icarus and EUHFORIA.}
         \label{fig:sw_map}
   \end{figure}

          \begin{figure}
   \centering
   \includegraphics[scale=0.30]{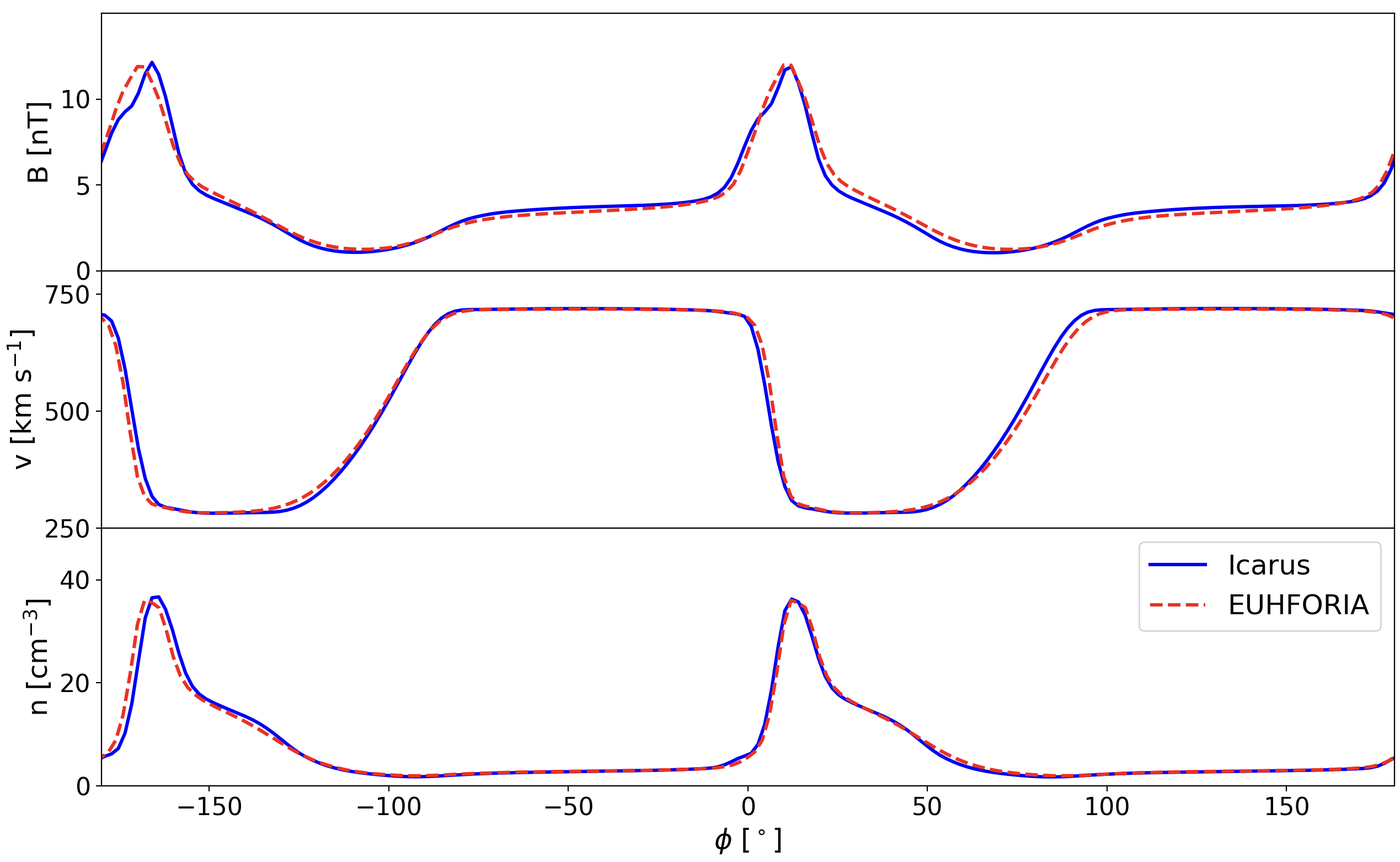}
      \caption{Plots of the magnetic field magnitude (top panel), the solar wind speed (middle panel), and the solar wind particle number density (bottom panel) versus longitude $\phi$ at radial distance $r = 1$~au and colatitude $\theta = 90^\circ$ obtained by Icarus (blue solid line) and EUHFORIA (red dashed line).}
         \label{fig:lon_profiles}
   \end{figure}

   \subsection{Simulation setup}\label{subsec:setup}

   For the simulations outlined in Sec.~\ref{sec:results} we use the following setup. The background solar wind of Icarus is displayed in the left panel of Fig.~\ref{fig:vr_comparison} and was obtained by using a synthetic solar wind map at the inlet boundary, 0.1~au, for the heliospheric wind model. The map is displayed in Fig.~\ref{fig:sw_map} and shows the radial solar wind velocity component of a tilted slow solar wind stream embedded in a fast wind stream which is based on the work in \citet{Pizzo-1991}. The band of slow solar wind represents a wind coming from a helmet streamer around the magnetic equator of the Sun. Such a configuration produces two CIRs as shown in Fig.~\ref{fig:vr_comparison}. Following \citet{Pizzo-1991}, in all simulations it is assumed that the magnetic axis of the Sun is tilted by $30^\circ$ with respect to its rotation axis.

   To determine the injection region for the test particles, we compute the divergence of the solar wind velocity of the heliospheric wind. As an example, in the right panel of Fig.~\ref{fig:vr_comparison}, we present a cross-section at $\theta = 90^\circ$ of the solar wind velocity divergence in the Icarus wind. The forward and reverse shocks of the CIR are clearly identifiable where the velocity divergence is negative. Given the perfect symmetry of this structure, we arbitrarily select the the CIR that extends to the lower half-plane of the right panel of Fig.~\ref{fig:vr_comparison} for injecting 1~MeV protons within a region spanning from 0.12 to 2~au, and limited in colatitude in the interval $90^\circ \pm 2^\circ$ at the shock wave. 
   We note that an injection energy on the order of 1~MeV is significantly higher than the anticipated energies of seed particle populations at CIR shocks. This divergence arises from the primary objective of our present study, which is to validate our model and illustrate the impact of AMR on acceleration efficiency within the model, rather than  replicating a real CIR event.

   The particles are injected according to the differential intensity formula, represented as
    \begin{align}
        j(r,E) = C\,\delta(E - 1\,\mathrm{MeV})\,r^{-2}\label{eq:differential_intensity}
    \end{align}
   with $C$ being a proportionality constant and $\delta$ the Dirac delta distribution, and where $\mathrm{div}\,\vect{V}_\mathrm{SW} <0$ and $j = p^2\,f$. In practical terms, this is accomplished within PARADISE by uniformly distributing 1~MeV protons throughout the region where $\mathrm{div}\,\vect{V}_\mathrm{SW} <0$ (tracing the CIR shock) and assigning them a statistical weight of $1/r^2$. To trace the shock, we used a modified version of the shock tracer code by \citet{Wijsen-etaL-2022}. Furthermore, we assume that $j$ is independent of time. To achieve this, we first calculate the Green's function solution for the FTE, which emerges when all particles are injected at time $t = 0$. Subsequently, we derive the steady-state solution for particle injection with constant time dependence by convolving the Green's function solution with a time-independent function. This is done for all simulations in Sec.~\ref{sec:results}.
   
   For the application of AMR in Sec.~\ref{subsec:amr_runs}, we use a region criterion to determine the refinement location, constraining the injection region to one CIR. Following the approach outlined in \citet{Wijsen-etal-2021} and \citet{Verbeke-etal-2022}, we estimate the longitudinal position $\hat{\phi}$ of the CIR that we wish to refine by
   \begin{align}
       \hat{\phi} = \phi + \frac{r - r_\mathrm{B}}{U}\,\Omega\,.\label{eq:region_criterion}
   \end{align}
   In Eq.~\eqref{eq:region_criterion}, $[r,\phi]$ denote coordinates in the domain, $r_\mathrm{B}$ is the radial distance at the inner boundary (i.e., 0.1~au), $U$ is the characteristic speed of the fast stream (in the present work, $U = 500\;$km\,s$^{-1}$), and $\Omega$ is the solar synodic rotation rate ($\Omega \approx 2.67 \cdot 10^{-6}$\,s$^{-1}$). To approximate the CIR spiral, Eq.~\eqref{eq:region_criterion} must satisfy the condition
   \begin{align}
       \phi_\mathrm{lower} < \hat{\phi} < \phi_\mathrm{upper},
   \end{align}
   with $\phi_\mathrm{lower}$ and $\phi_\mathrm{upper}$ denoting the upper and lower longitudinal limits, respectively. We note that the standard setting of Icarus uses the solar sidereal rotation rate, but since EUHFORIA uses the solar synodic rotation rate, we adjusted it for the present study.

   The MPI-AMRVAC framework contains a number of numerical schemes and shock capturing methods that can be selected. In this particular study, we choose the second-order (in time and space) total variation diminishing Lax-Friedrichs (TVDLF) scheme as our shock-capturing method\citep{Toth-Odstrcil-1996} combined with the Woodward limiter \citep{Woodward-Colella-1984}. The TVDLF is a robust scheme, but more diffusive in comparison to other solvers, such as the Harten-Lax-van Leer (HLL) Riemann solver. 
   A key distinction between EUHFORIA and Icarus is that in EUHFORIA $\nabla \cdot \textbf{\textit{B}}$ is prescribed to be 0. While MPI-AMRVAC has the option to also use a constraint transport approach, in Icarus, small non-zero values of $\nabla \cdot \textbf{\textit{B}}$ are locally allowed. The parabolic cleaning method by \citet{Keppens-etal-2003} is included, which diffuses any local non-zero divergence of the magnetic field.

    To validate the Icarus$+$PARADISE model in Sec.~\ref{subsec:validation}, we conducted simulations replicating outcomes of the EUHFORIA$+$PARADISE model. To achieve this, we generated a similar heliospheric wind in EUHFORIA, employing the same coronal map utilised for Icarus, as presented in Fig.~\ref{fig:sw_map}. In Fig.~\ref{fig:lon_profiles}, we provide a comparison of the longitudinal profiles of the magnetic field magnitude (top panel), the solar wind speed (middle panel), and the particle number density (bottom panel) at 1.0~au and $\theta = 90^\circ$. The results are in reasonable agreement with slight differences in all three solar wind parameters which are due to the entirely different numerical schemes used in EUHFORIA and Icarus. The spatial domain of both winds covers a radial distance of 0.1~au to 2~au, since particle acceleration in CIRs often occurs beyond 1.0~au, and a colatitudinal and longitudinal range of $30^\circ$ to $150^\circ$ and $0^\circ$ to $360^\circ$, respectively. The resolution of both winds is the same with 600 cells in radial direction or $\Delta r \approx 1.2$~$R_\mathrm{S}$, 60 cells in colatitude and 180 cells in longitude, which corresponds to an angular resolution of $\Delta \theta = \Delta \phi = 2^\circ$.

    The simulations for the comparison of different levels of AMR in Sec.~\ref{subsec:amr_runs} utilise the same Icarus solar wind as in Sec.~\ref{subsec:validation}. However, on AMR level 1, which corresponds to a uniform initial grid, we choose a low resolution of only 300 cells in radial direction, 40 cells in colatitude and 96 cells in longitude. With each subsequent AMR level, the number of cells in each direction doubles, at least locally, i.e., where the given AMR criteria are satisfied. In the highest AMR level achievable in the chosen simulation setup, AMR level 5, there are 'effectively' 4800 cells in the radial direction, 640 'effective' cells in colatitude, and 1536 'effective' cells in longitude, reflecting the highest refinement level.

    Finally, we note that we made two modifications in the inner boundary conditions in comparison to the standard Icarus version available in MPI-AMRVAC. Since larger discrepancies in $v_\phi$ and $B_\phi$ in Icarus and EUHFORIA at and near the inner boundary were noticed resulting from the use of different coordinate systems, we changed the inner boundary in Icarus to match those in EUHFORIA. The longitudinal velocity component at the inner boundary was altered to $v_\phi = - r_\mathrm{B}\,\Omega\,\sin(\theta)$ in order to obtain $v_\phi \approx 0$ at the inner boundary in the inertial frame. According to \citet{Pomoell-Poedts-2018}, we also adjusted the longitudinal magnetic field component to $B_\phi = B_\mathrm{r}\,v_\phi / v_\mathrm{r}$, where $B_\mathrm{r}$ and $v_\mathrm{r}$ are the radial magnetic and velocity component, respectively. This change in $B_\phi$ was made to ensure that the electric field is zero in the corotating frame, a necessary condition for obtaining a steady-state solution of the solar wind in the corotating frame, as implemented in EUHFORIA.

          \begin{figure}[h!]
   \centering
   \includegraphics[scale=0.35]{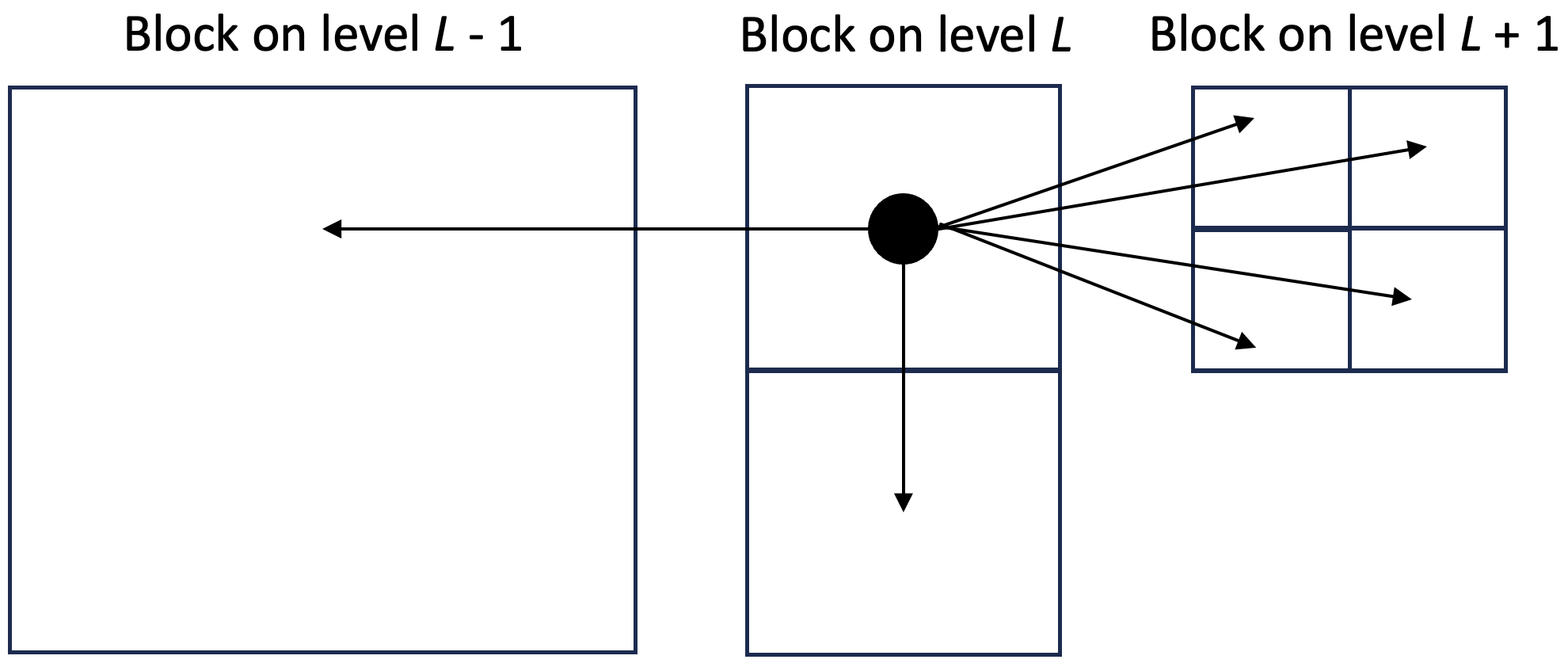}
      \caption{Illustration of the AMR nesting and communication between blocks. Using AMR likely requires blocks of some level $L$ to communicate both to coarse blocks (level $L-1$) and further refined blocks (level $L+1$).}
         \label{fig:amr_illustration}
   \end{figure}

   \subsection{Coupling of PARADISE to Icarus}\label{subsec:coupling_PARADISE_Icarus}

The PARADISE particle transport code employs a grid-free numerical approach for the propagation of energetic test particles. However, the input background solar wind configurations required by PARADISE are provided on a numerical grid, containing parameters of the solar wind plasma. To integrate the set of equations constituting the FTE for a specific particle, it becomes necessary to locate the test particle on the numerical grid of the particular heliospheric solar wind model, read out the solar wind plasma variables of the grid points surrounding the particle, and interpolate them at each computational step. For this purpose, the method of trilinear interpolation that has been applied for the previously coupled EUHFORIA model has been adopted also for the implementation of the Icarus model due to the rectangular arrangement of solar wind variables within Icarus.

While EUHFORIA utilises a uniform and equidistant grid contained in a single large rectangular block, the Icarus grid adopts a more complex structure comprising multiple blocks. In the case of Icarus, each level of refinement divides the target block into halves along each coordinate direction. As higher levels of AMR are applied, these blocks with different AMR levels follow a strict nesting protocol, that is, two blocks touching each other either orthogonally or diagonally, will never exhibit an AMR level difference exceeding one \citep{Nool-Keppens-2002}.

The primary challenge encountered when coupling PARADISE with Icarus arose from the block-based structure of the Icarus grid when multiple levels of AMR are applied. When a particle is close to the boundary of a block, that is, when it is in some form outside the internal grid structure of the current block, communication between multiple neighbouring blocks becomes necessary to provide requisite data for numerical computation processes such as trilinear interpolation or finite volume differencing. This issue is illustrated in Fig.~\ref{fig:amr_illustration} that has been adapted from the MPI-AMRVAC documentation where further information about the AMR implementation can be found\footnote{\url{https://amrvac.org/md_doc_amrstructure.html}}. In cases involving more than two levels of AMR, it becomes evident that a block of some level $L$ must eventually communicate with either a so-called parent block of level $L - 1$, or a so-called children block of level $L + 1$. Communication between blocks of the same level occurs both with and without applied AMR.

The solution of having ghost cells for each block, as employed by the built-in particle tracer in AMRVAC, was not feasible for PARADISE due to resulting significantly larger data files storing these ghost cell values. As an alternative, we chose to individually address the various scenarios based on a particle's position relative to the numerical grid and acquire all necessary data from neighbouring blocks as needed.

\section{Simulation results}\label{sec:results}

In this section, we showcase simulation results obtained from the Icarus$+$PARADISE model. Implementing the setup described in Sec.~\ref{subsec:setup}, we first validate our new model by reproducing results of the previously established EUHFORIA$+$PARADISE model using in Icarus a uniform grid (Sec.~\ref{subsec:validation}). Subsequently, we explore the impact of higher levels of AMR to illustrate its effects (Sec.~\ref{subsec:amr_runs}).

\subsection{Validation}\label{subsec:validation}

\begin{figure*}[t!]
    \centering
    \subfloat{{\includegraphics[scale=0.34]{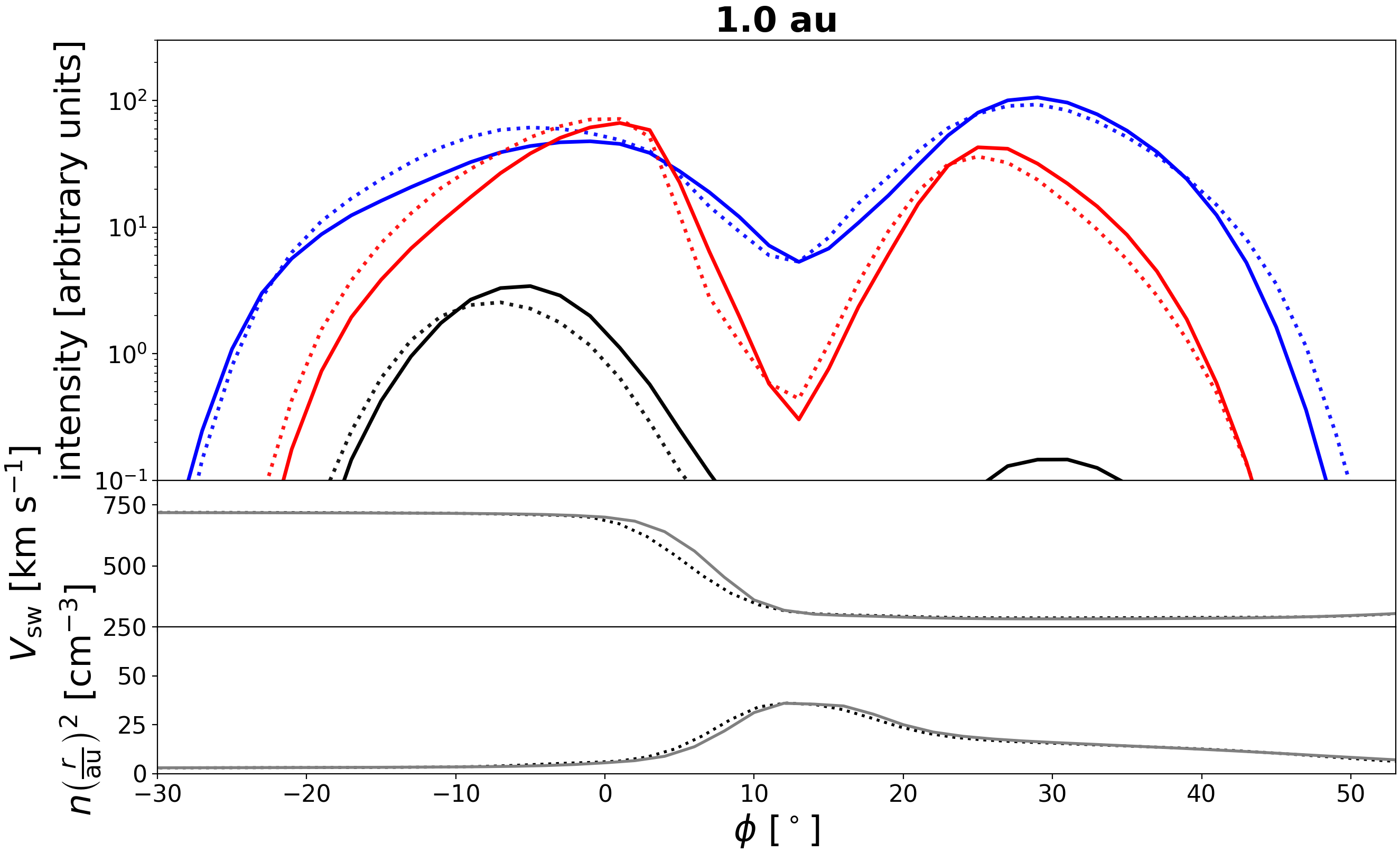}}}%
    
    \subfloat{{\includegraphics[scale=0.32]{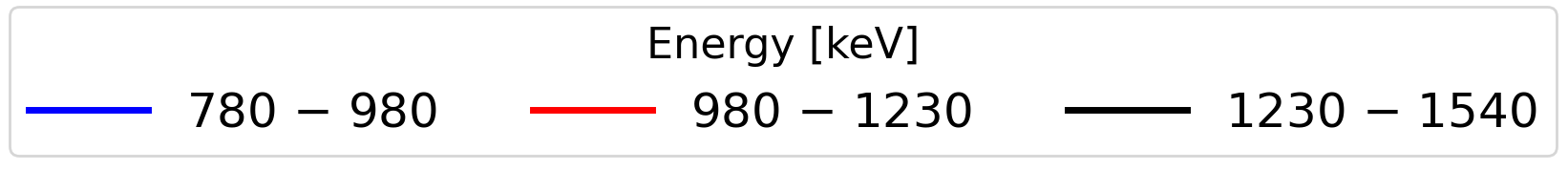}}}%
    
    \caption{Longitudinal profiles of intensities (top panel) obtained for 1~MeV protons at $\theta = 90^\circ$ and $r = 1.0$~au. The intensities obtained in the Icarus wind are represented by dotted lines, while the EUHFORIA-based trends are depicted by solid lines, each in their respective colours. Similarly, the solar wind speed and solar wind particle number density (middle and bottom panel, respectively) are shown in dotted lines for Icarus and solid lines for EUHFORIA.}%
    \label{fig:validation_azimuthal_profiles_10au}%
\end{figure*}

\begin{figure*}[t!]
    \centering
    \subfloat{{\includegraphics[scale=0.34]{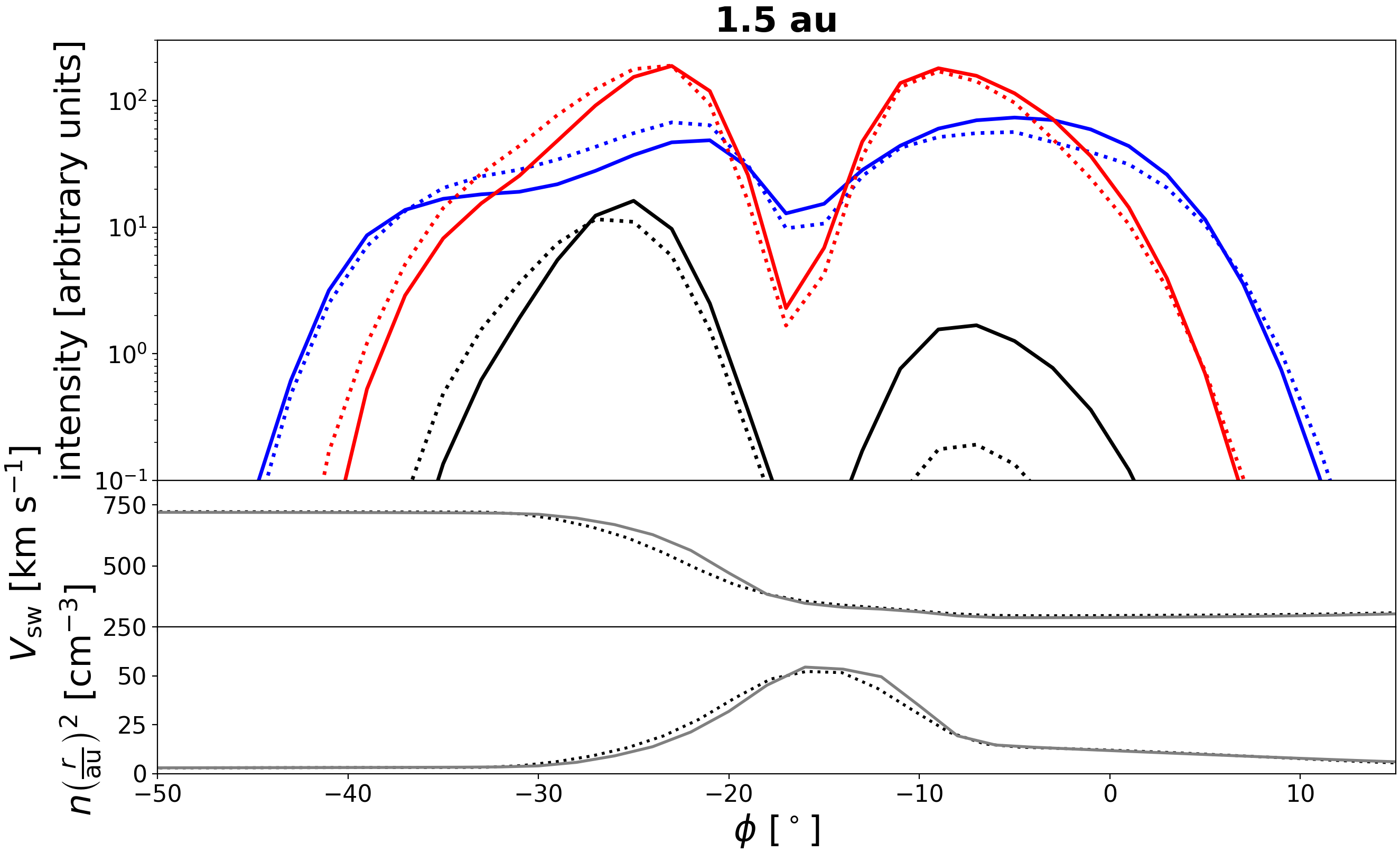}}}%
    
    \subfloat{{\includegraphics[scale=0.32]{new_images/valid_legend.png}}}%
    
    \caption{Longitudinal profiles of intensities (top panel) obtained for 1~MeV protons at $\theta = 90^\circ$ and $r = 1.5$~au. The intensities obtained in the Icarus wind are represented by dotted lines, while the EUHFORIA-based trends are depicted by solid lines, each in their respective colours. Similarly, the solar wind speed and solar wind particle number density (middle and bottom panel, respectively) are shown in dotted lines for Icarus and solid lines for EUHFORIA.}%
    \label{fig:validation_azimuthal_profiles_15au}%
\end{figure*}

Figures~\ref{fig:validation_azimuthal_profiles_10au} and \ref{fig:validation_azimuthal_profiles_15au} show longitudinal profiles of the particle intensities obtained by PARADISE at radial distances $r = 1$~au (Fig.~\ref{fig:validation_azimuthal_profiles_10au}) and $r = 1.5$~au (Fig.~\ref{fig:validation_azimuthal_profiles_15au}). In each figure, the top panel shows the intensity profiles of three selected energy channels that are defined in the legend below the figures. Additionally, we provide the longitudinal profiles of solar wind speed (middle panel) and the particle number density (bottom panel). The Icarus-based results are shown in dotted lines, while the results obtained in the EUHFORIA wind are shown in the same figures in the same colours, but in solid lines. Similarly, the solar wind parameter profiles are shown in dotted format for Icarus and in solid for EUHFORIA. 

For both models, we observe the result of adiabatic deceleration (channel below 1~MeV) and shock acceleration (channels above 1~MeV). Furthermore, the intensity profiles reveal a characteristic two-peaked structure typically associated with a CIR. The right intensity peak corresponds to particles accelerated at the forward shock of the CIR, whereas the valley between the two peaks coincides with the stream interface (SI) of the CIR, that is, the transition between the compressed slow and fast wind. The left intensity peak is attributed to particles originating from the reverse shock.

In the two lowest energy channels, the two models exhibit good agreement, resembling in shape and following a similar trend. It can be noted that, depending on the longitude, the Icarus-based intensities exceed the EUHFORIA-based results, and vice versa. These small differences can be attributed to the small differences between the underlying solar wind configurations. This variability is expected, given that the underlying models are not identical, and energetic particles traverse a significant distance through these two solar winds before reaching observers. Cumulatively, small differences in both solar winds can impact the particles' properties. Moreover, in the highest energy channel, it can be seen that the intensities in EUFHORIA peak higher than in Icarus. This difference can be attributed to the fact that, at the same resolution, Icarus exhibits a slightly higher level of diffusion compared to EUHFORIA. Consequently, the shock in Icarus is less steep than in the EUHFORIA solar wind, resulting in a reduction of particle acceleration efficiency. However, as demonstrated in the following section, the implementation of AMR enables Icarus to achieve a resolution at the shock significantly surpassing the one depicted in Fig.~\ref{fig:validation_azimuthal_profiles_10au} and \ref{fig:validation_azimuthal_profiles_15au}.

\subsection{Effect of AMR application}\label{subsec:amr_runs}

\begin{figure*}[t!]
\begin{subfigure}{0.48\textwidth}
\includegraphics[width=\linewidth]{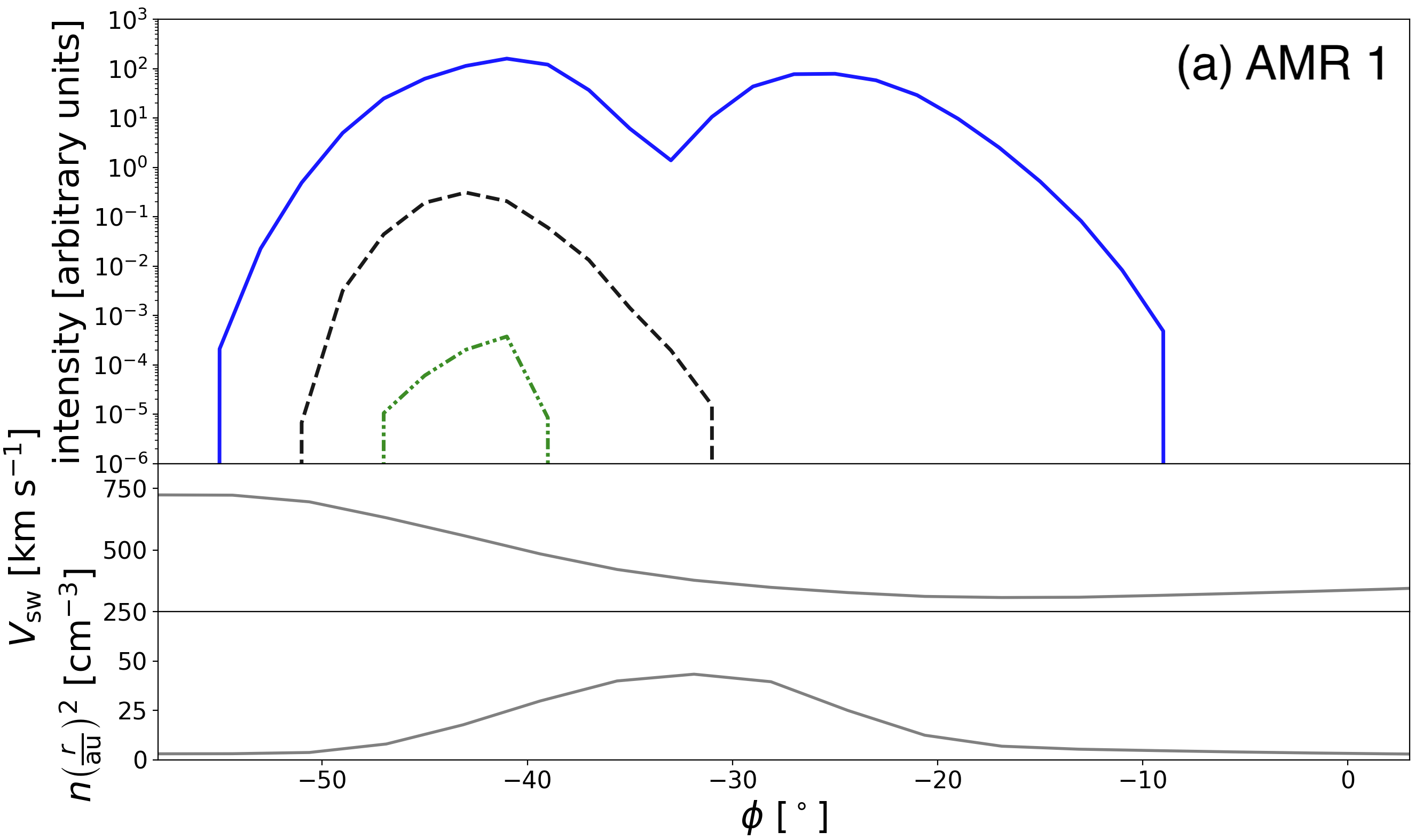}
\end{subfigure}\hspace*{\fill}
\begin{subfigure}{0.48\textwidth}
\includegraphics[width=\linewidth]{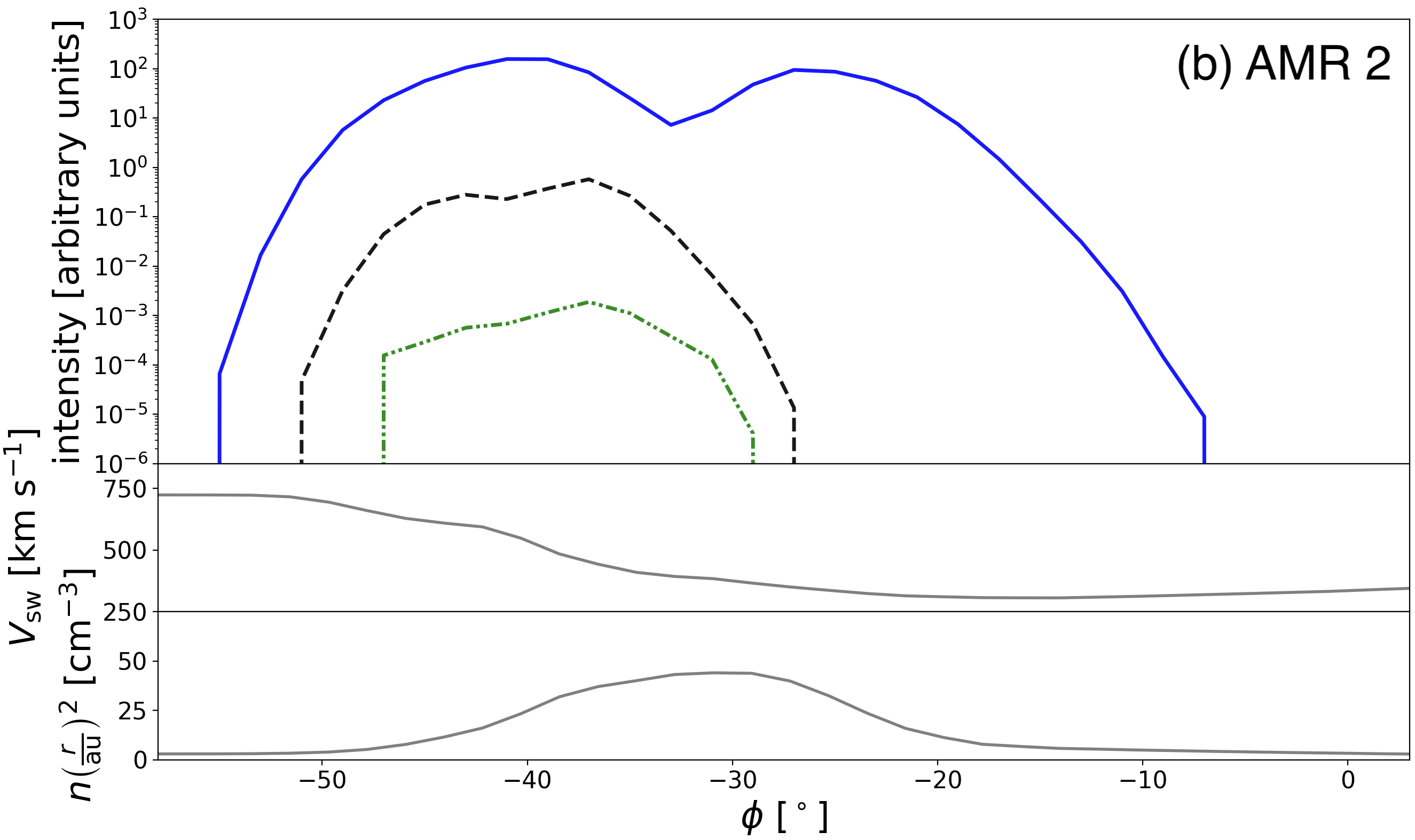}
\end{subfigure}

\medskip
\begin{subfigure}{0.48\textwidth}
\includegraphics[width=\linewidth]{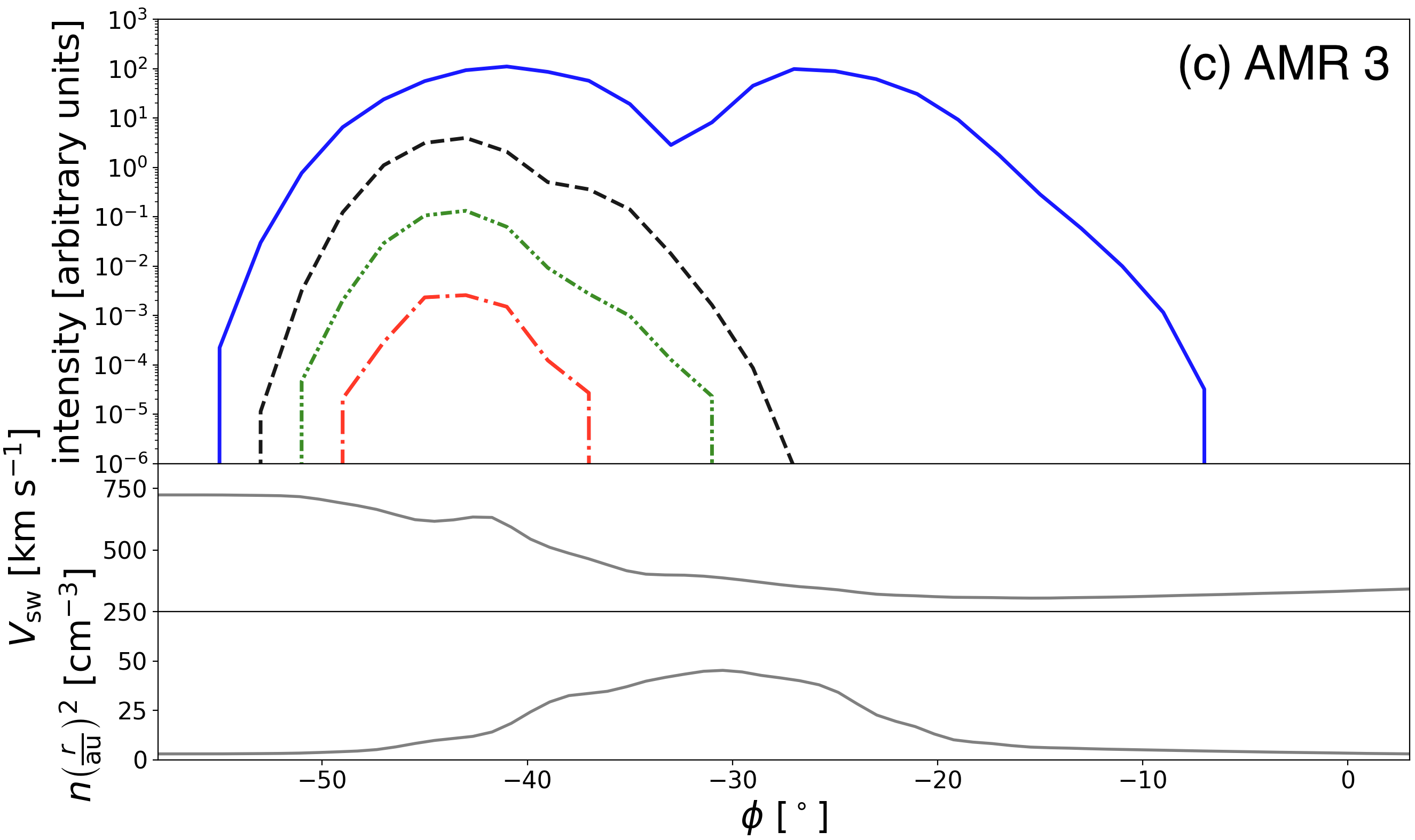}
\end{subfigure}\hspace*{\fill}
\begin{subfigure}{0.48\textwidth}
\includegraphics[width=\linewidth]{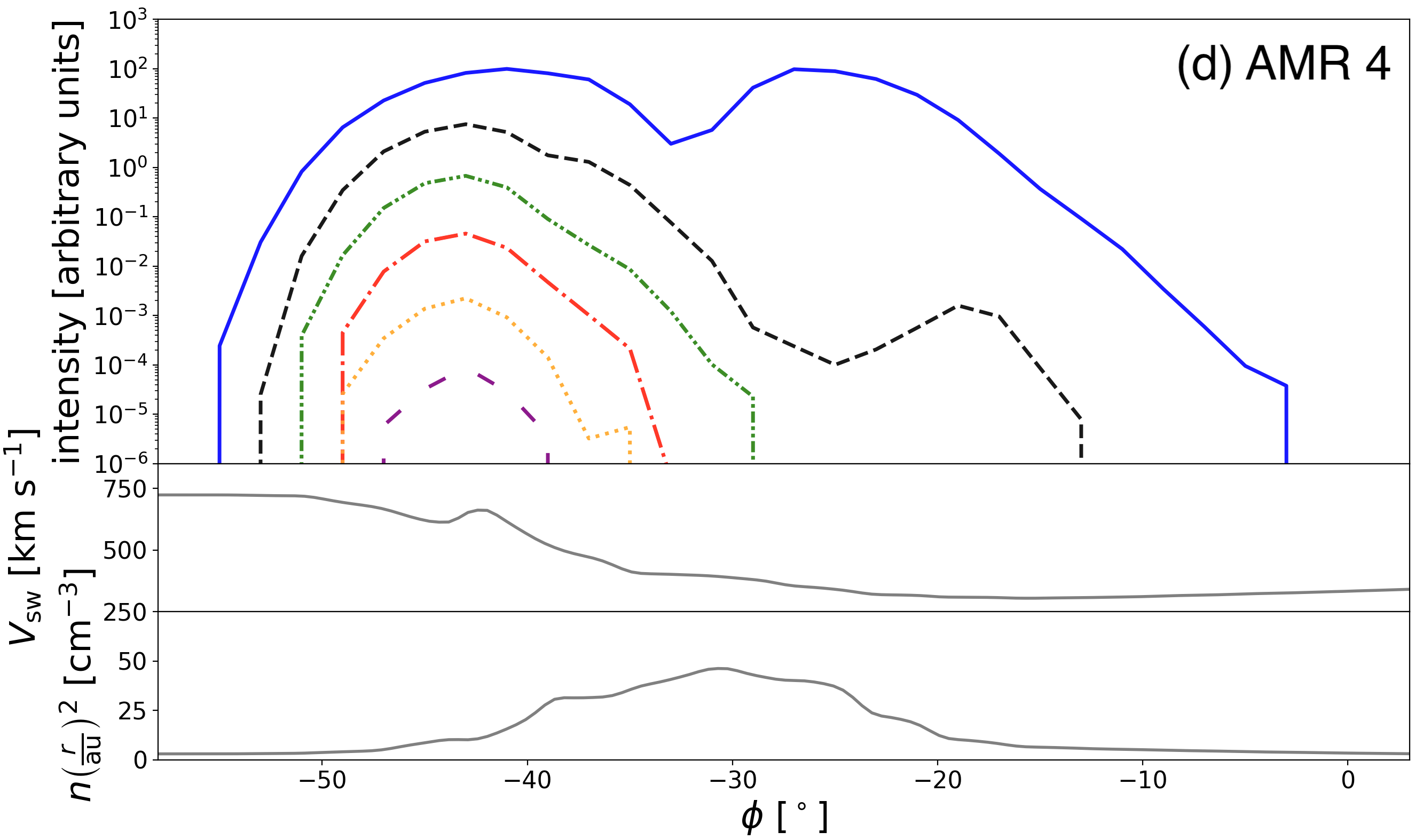}
\end{subfigure}

\medskip
\begin{subfigure}{0.48\textwidth}
\includegraphics[width=\linewidth]{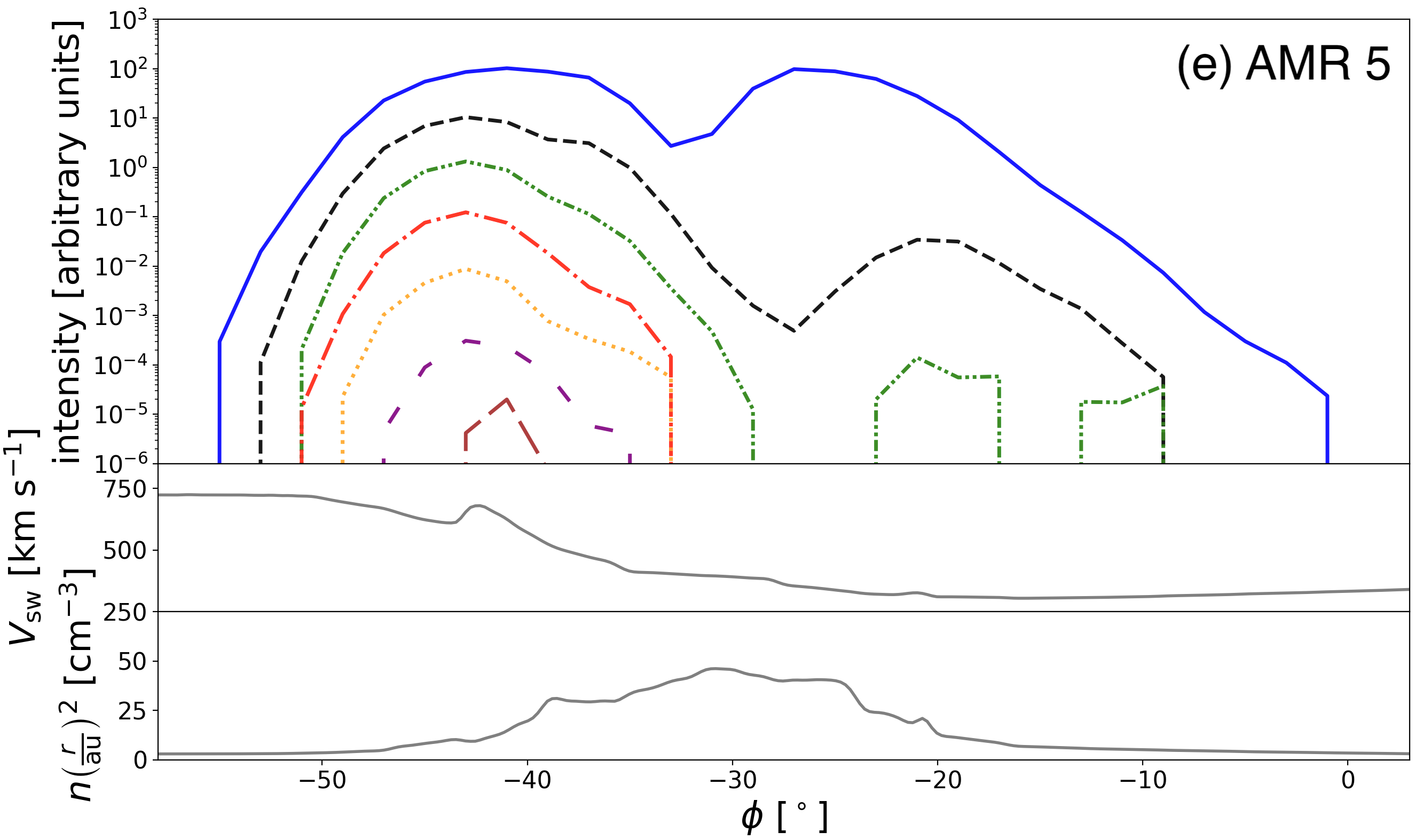}
\end{subfigure}\hspace*{\fill}
\begin{subfigure}{0.48\textwidth}
\includegraphics[width=\linewidth]{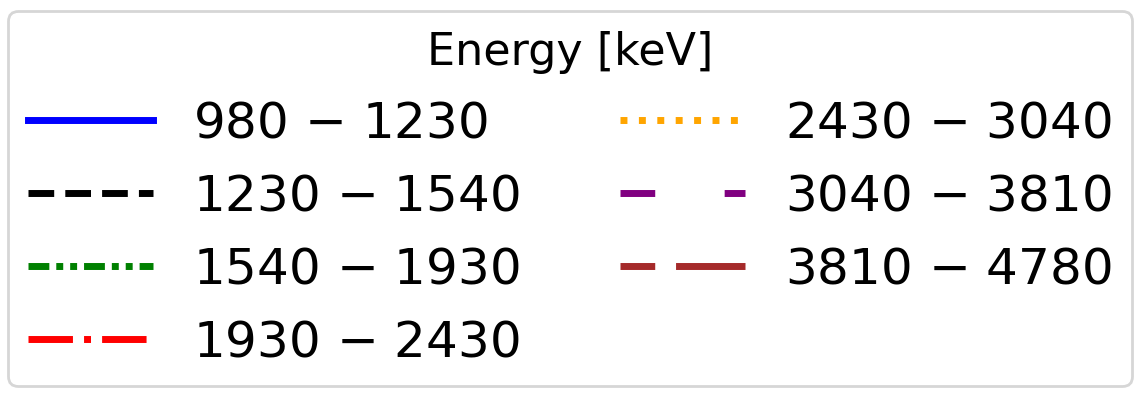}
\end{subfigure}

\caption{Longitudinal profiles of intensities obtained for 1 MeV protons in the Icarus wind at $\theta = 90^\circ$ for five levels of AMR (panels a through e). Each plot displays intensities at $r = 1.8$~au together with the background solar wind speed (middle panels) and solar wind particle number density (bottom panels).} \label{fig:amr_plots}
\end{figure*}

In contrast to similar heliospheric MHD models such as ENLIL or EUHFORIA, which employ uniform equidistant grids, the MPI-AMRVAC framework empowers Icarus to utilise AMR to refine regions of interest while maintaining a low resolution in other regions. We illustrate the application of five levels of AMR in the panels of Fig.~\ref{fig:amr_plots} that are similarly structured as Figs.~\ref{fig:validation_azimuthal_profiles_10au} and \ref{fig:validation_azimuthal_profiles_15au} with the longitudinal intensity profiles in the top panels, the solar wind speed profiles in the middle panels, and the solar wind particle number density in the bottom panels in each subfigure. Here, AMR level 1 corresponds to a uniform grid with the resolution as described in Sec.~\ref{subsec:setup}. The subfigures in Fig.~\ref{fig:amr_plots} depict the intensities at $r = 1.8$~au. 

Similarly to Figs.~\ref{fig:validation_azimuthal_profiles_10au} and \ref{fig:validation_azimuthal_profiles_15au}, we observe two peaks in the intensities, the right peak corresponding to the forward shock, and the left peak corresponding to the reverse shock. For the higher energy channels, the intensity peak associated to the forward shock becomes only visible by increasing the resolution (i.e., higher AMR level).  With increasing level of AMR, the channels of higher energies, that is the channels measuring the accelerated particles, are increasingly populated. It is crucial to note that shock waves simulated by 3D MHD models are typically much wider than actual shock waves in interplanetary space. Since the ratio of the MFP length to the width of the shock plays a pivotal role, simulations with a more refined shock (here, due to higher levels of AMR) result in more efficient particle acceleration for the same MFP length, see e.g., \citet{Wijsen-etaL-2022}.

\begin{figure*}
\centering
\subfloat{\includegraphics[scale=0.17]{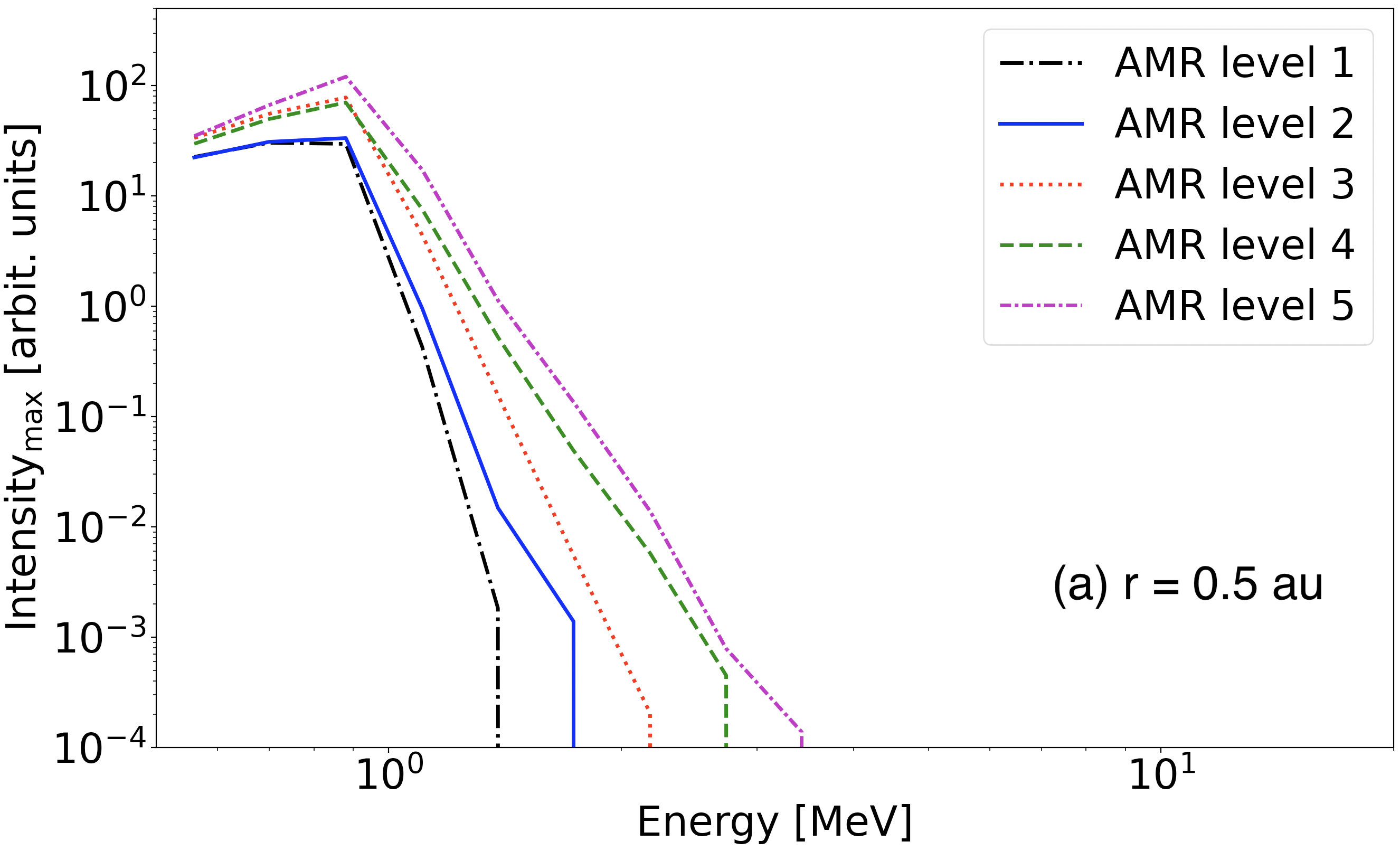}}\hfil
\subfloat{\includegraphics[scale=0.17]{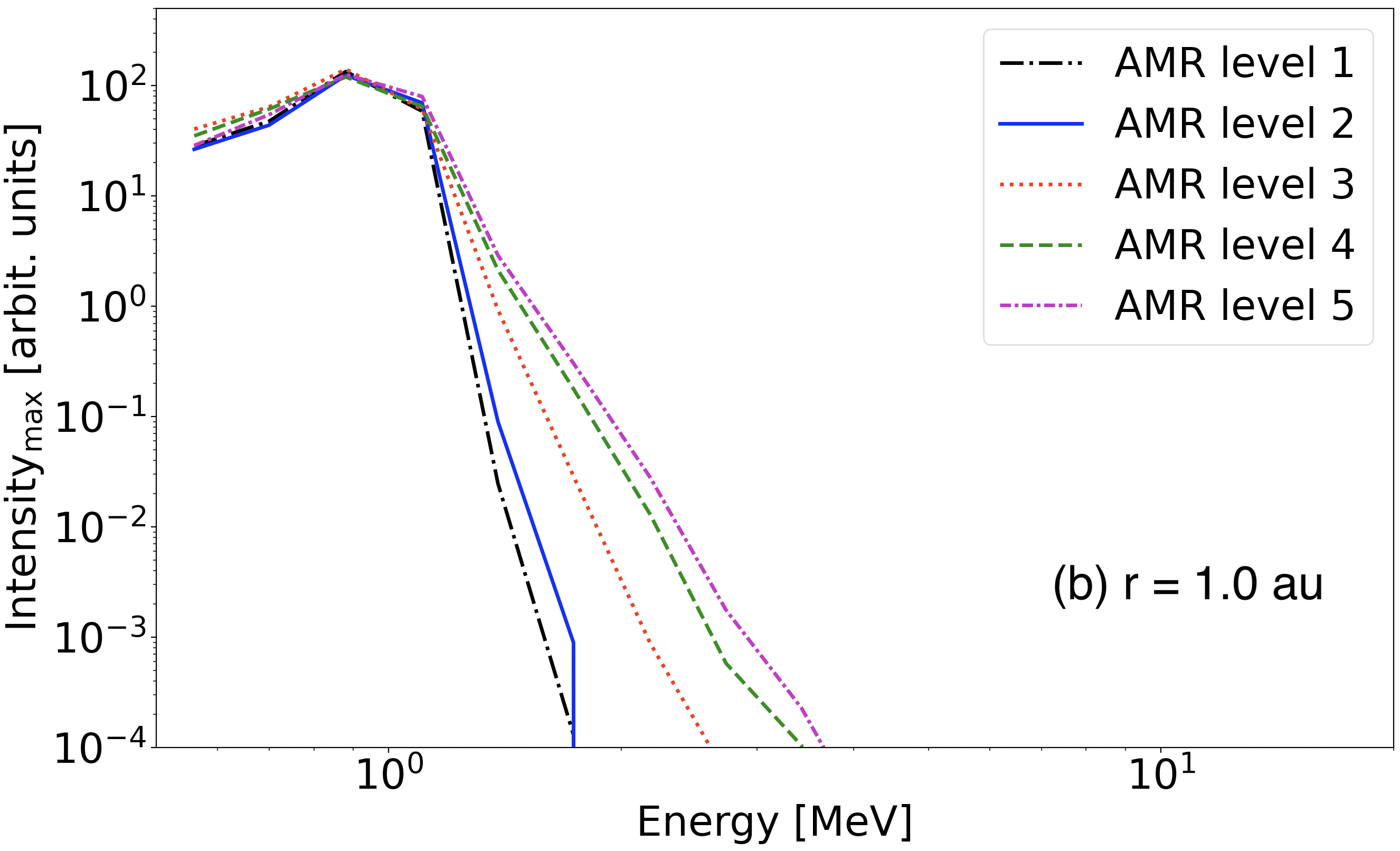}}

\subfloat{\includegraphics[scale=0.17]{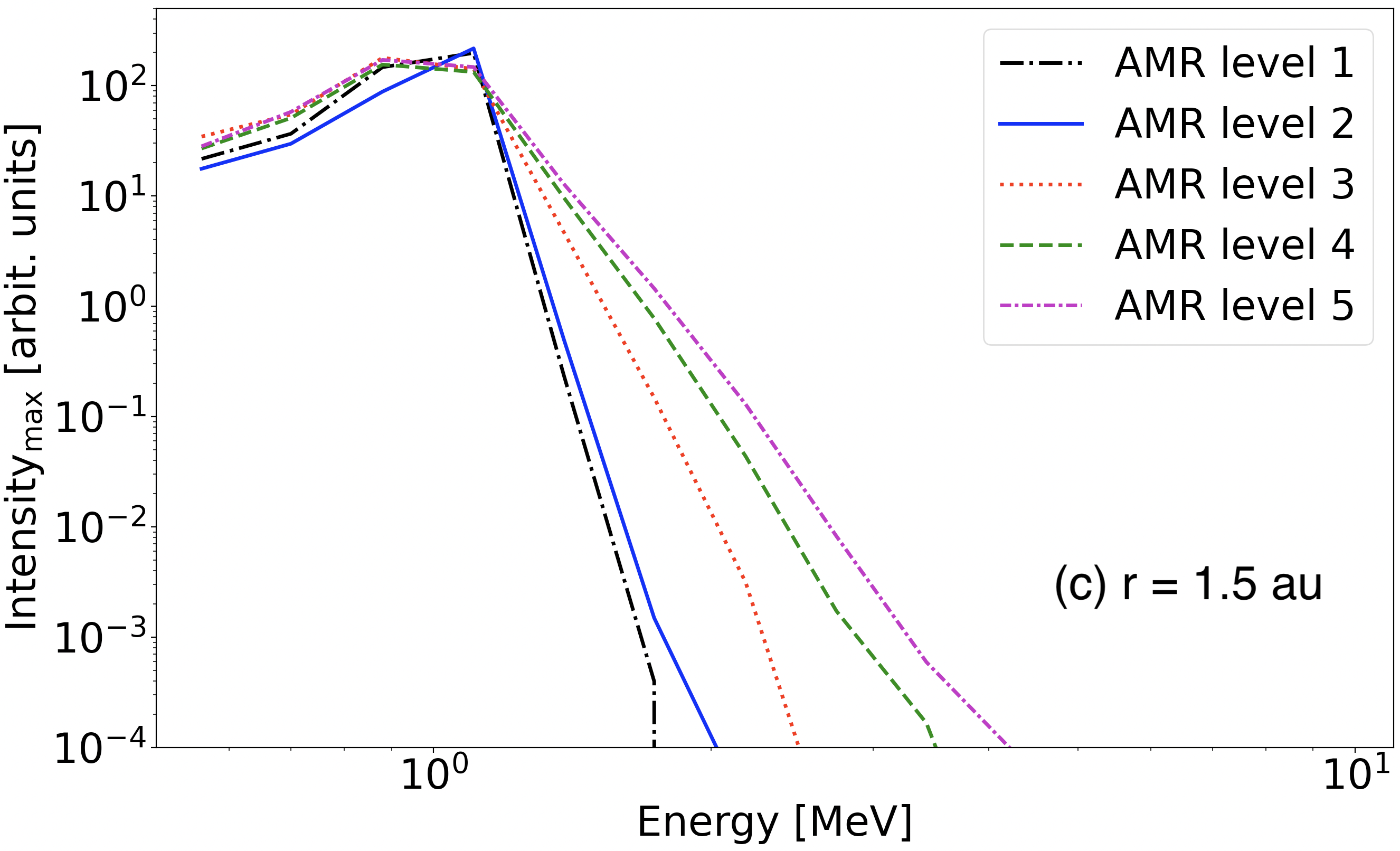}}\hfil 
\subfloat{\includegraphics[scale=0.17]{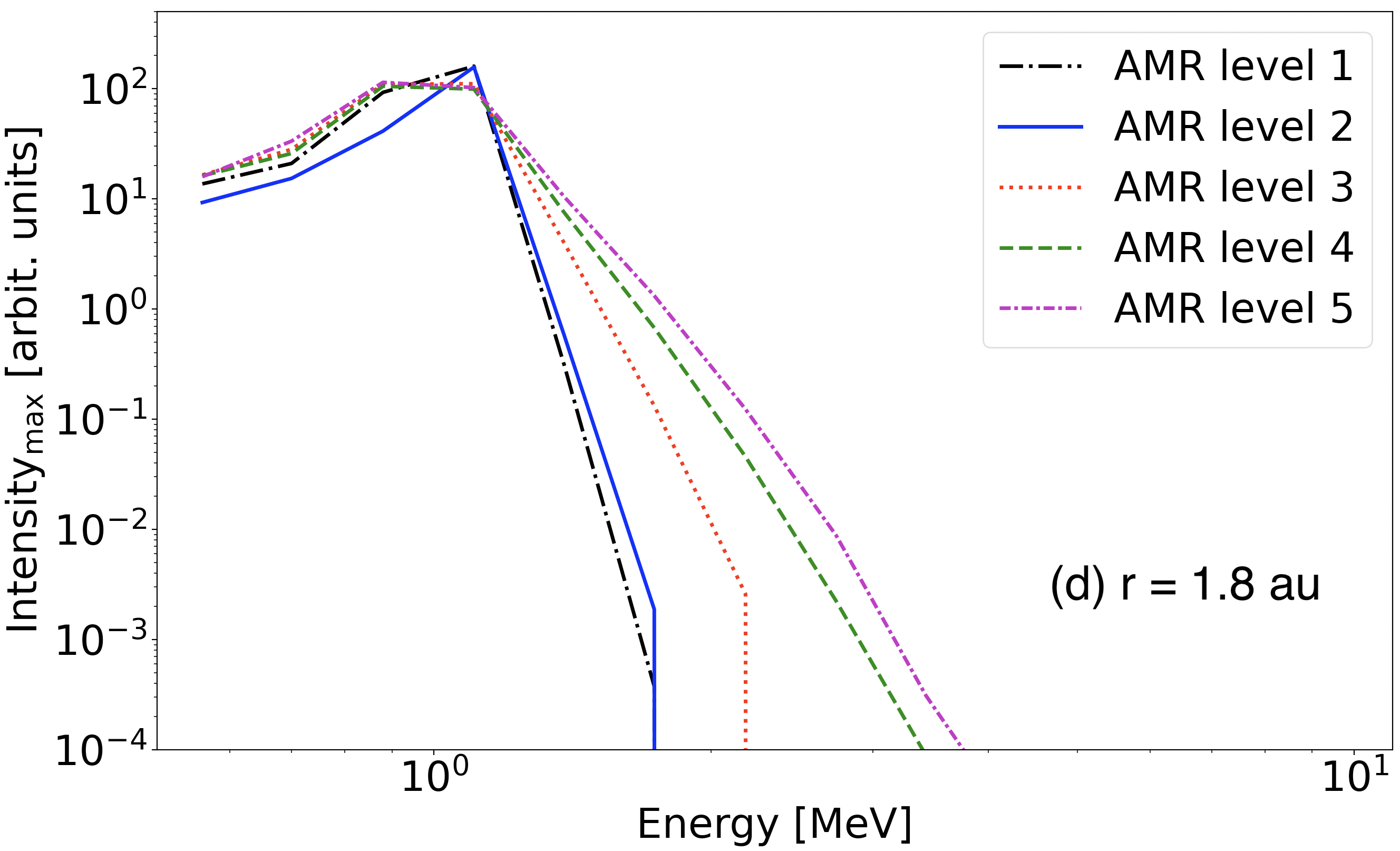}}
\caption{Plots of intensities versus energy at $r = 0.5$~au (panel a), $r = 1.0$~au (panel b), $r = 1.5$~au (panel c), and $r = 1.8$~au (panel d), using different levels of AMR. The panels show the intensity peaks of the different energy channels at the reverse shock from a simulation up to 2~au.}\label{fig:intensity_energy_rs_2au}
\end{figure*}

To elucidate the impact of higher levels of AMR on particle acceleration, Fig.~\ref{fig:intensity_energy_rs_2au} displays intensity-energy profiles at different radial distances at the reverse shock. The intensity curves consist of the peaks of the corresponding energy channel at the particular radial distance, where panel (a) show the results at 0.5~au, panel (b) at 1~au, panel (c) at 1.5~au, and panel (d) at 1.8~au. All panels consistently indicate an increase in the intensity at higher energies with increasing level of AMR. 

With increasing resolution, the particle intensities and their energy spectra should converge if the underlying solar wind converges. However, it should be kept in mind that by increasing the resolution of the MHD simulation, we are resolving smaller-scaled solar wind structures, which can subsequently affect the energetic particles. While the results in Fig.~\ref{fig:intensity_energy_rs_2au} do not show convergence yet, the differences between consecutive AMR levels can be seen to progressively decrease.

\section{Summary and discussion}\label{sec:summary_conclusions}

    The significance of energetic particles generated in the heliosphere is growing due to humanity's dependence on space technology and its renewed interest in space exploration. Consequently, there is a pressing need for the development of numerical tools to reliably and timely predict harsh events. In order to make progress in our space weather forecasting capabilities and our theoretical understanding of transport and acceleration of solar energetic particles, we have introduced the Icarus$+$PARADISE model. This two-part-model comprises the 3D MHD code Icarus, integrated into the MPI-AMRVAC framework, which generates solar wind background configurations. Accompanying Icarus is the energetic particle transport code PARADISE, which takes the Icarus wind as input to propagate energetic particles as test particles through it and tracks their acceleration and deceleration by solving the FTE in a stochastic manner.

    To introduce the new model, we presented a theoretical study using a synthetic solar wind map for the coronal model providing Icarus with inner boundary conditions to generate a solar wind configuration that contains two CIRs. As part of the validation process, we reproduced outcomes from the earlier EUHFORIA$+$PARADISE model and showcased longitudinal intensity profiles at different radial distances. The simulation results demonstrated overall good agreement, exhibiting a strong similarity in the lower energy channels. However, some disparities surfaced in the higher channels, where the intensities derived from EUHFORIA surpassed those obtained in the Icarus simulation for the same underlying MHD resolution. We attribute these variations in intensity profiles to differences between the two solar wind models, stemming from the distinct numerical schemes employed by EUHFORIA and Icarus. For instance, the TVDLF scheme used in Icarus for shock capturing, was found to be more diffusive compared to the one utilised in EUHFORIA. Alternative shock capturing schemes, such as the Harten-Lax-van Leer (HLL) Riemann solver or Roe-type approximate Riemann solvers, are available in the MPI-AMRVAC framework \citep{Keppens-etal-2023} and are known to be less diffusive than the TVDLF scheme. They will be explored for evaluation in future studies due to their potential advantages.

    The MPI-AMRVAC framework employed by Icarus offers the flexibility to include AMR. In our study, we strategically utilised a region criterion to refine only the shock region where we injected the test particles, while maintaining a lower resolution for the remaining solar wind. Using five levels of AMR, we displayed longitudinal profiles of the intensities at 1.8~au. Furthermore, we extracted the intensity peaks for the different energy channels and illustrated the corresponding intensity-energy plots at the reverse shock at different radial distances. Our findings consistently demonstrated that as the level of AMR increased, a more pronounced particle acceleration was observed. This result holds significant implications for models that simulate particle acceleration at shock waves generated by MHD models, see for instance, \citet{Wijsen-etaL-2022} using EUHFORIA+PARADISE to study proton acceleration in the inner heliosphere, or \citet{Young-etal-2021} using EPREM and the CORona-HELiosphere (CORHEL) model to investigate proton acceleration in the corona.

    In our future work, we anticipate further applications of the Icarus$+$PARADISE model. These potential applications encompass, among others, the examination of solar wind configurations of real events involving CMEs and observed SEP events, mirroring previous studies with EUHFORIA$+$PARADISE. Furthermore, Icarus is also capable to implement grid stretching, that is the ability to keep grid cells in a more cubic shape that in an otherwise equidistant grid in spherical coordinates would become more elongated. Performance of Icarus with applied radial grid stretching with combination of AMR is demonstrated in \citet{Baratashvili-etal-2022}. Such an approach can be tested for SEP studies with sufficient AMR levels in order to gain significant speed-up in computation time, while obtaining high resolution data where needed. 
    Furthermore, a major update for Icarus$+$PARADISE is foreseen, for instance, a modification of PARADISE such to load only the necessary blocks of the Icarus snapshots, rather than loading entire data files. This and other enhancements are designed to accelerate simulations inside PARADISE, enable the use of higher levels of AMR, and augment the already significant acceleration of simulation time achieved by Icarus.

\begin{acknowledgements}
       The authors thank Rony Keppens and Fabio Bacchini from KU Leuven, Belgium, for fruitful discussions regarding the MPI-AMRVAC framework. In addition, the authors thank the anonymous referees for their helpful comments that improved the quality of the manuscript.
      E.H.\ is grateful to the Space Weather Awareness Training Network (SWATNet) funded by the European Union's Horizon 2020 research and innovation program under the Marie Skłodowska-Curie grant agreement No. 955620. Furthermore, E.H.\ acknowledges the travel grant V477923N from the Fonds voor Wetenschappelijk Onderzoek – Vlaanderen (FWO). 
      N.W.\ acknowledges support from the Research Foundation - Flanders (FWO-Vlaanderen, fellowship no.\ 1184319N). 
      R.V.\ acknowledges the support of the Research Council of Finland (grant no. 352847).
      We also acknowledge support from the European Space Agency project "Heliospheric modelling techniques“ (ESA Contract No. 4000133080/20/NL/CRS) as part of the GSTP (General Support Technology Programme), the projects C14/19/089  (C1 project Internal Funds KU Leuven), G.0B58.23N and G.0025.23N (WEAVE)   (FWO-Vlaanderen), 4000134474 (SIDC Data Exploitation, ESA Prodex-12), and Belspo project B2/191/P1/SWiM. 
      Computational resources and services used in this work were provided by the VSC (Flemish Supercomputer Center), funded by the Research Foundation - Flanders (FWO) and the Flemish Government  – department EWI.
\end{acknowledgements}

\bibliography{jswsc}
   

\end{document}